\documentclass[12pt]{iopart}

\usepackage{iopams}

\usepackage[utf8]{inputenc}
\usepackage{natbib}
\usepackage{setstack}
\usepackage[ngerman,english]{babel}
\usepackage{graphicx}
\usepackage[nooneline]{subfigure}
\usepackage{upgreek}
\usepackage{url}
\usepackage{xspace}
\usepackage{paralist}
\usepackage{verbatim}
\usepackage{icomma}

\newenvironment{align*}{\begin{eqnarray*}}{\end{eqnarray*}}

\newcommand{\bilder}{.}

\newcommand{\db}{0.31\textwidth}

\setcitestyle{comma,aysep={}}

\newcommand{\rs}{r_\mathrm{S}}
\newcommand{\rrmi}{r_\mathrm{m}}
\newcommand{\ud}{\mathrm{d}}

\begin{document}

\title[Sector models: III. Spacetime geodesics]{%
Sector models---A toolkit for teaching general relativity:
III. Spacetime geodesics
}

\author{U Kraus and C Zahn}

\bigskip
\address{Institut f\"ur Physik, Universität Hildesheim,
Universitätsplatz 1, 31141 Hildesheim, Germany}

\eads{
\mailto{ute.kraus@uni-hildesheim.de},
\mailto{corvin.zahn@uni-hildesheim.de}
}

\begin{abstract}

{\parindent=0pt

Sector models permit
a model-based approach to the
general theory of relativity.
The approach has its focus on the geometric concepts
and uses no more than elementary mathematics.
This contribution shows how to construct
the paths of light and free particles
on a spacetime sector model.
Radial paths close to a black hole are used by way of example.
We outline two workshops on gravitational redshift and
on vertical free fall, respectively,
that we teach for undergraduate students.
The workshop on redshift does not require knowledge
of special relativity; the workshop on particles in free fall
presumes familiarity with the Lorentz transformation.
The contribution also describes a simplified calculation
of the spacetime sector model that students can carry out
on their own if they are familiar with the Minkowski metric.
The teaching materials presented in this paper
are available online for teaching purposes at
{\tt www.spacetimetravel.org}.


}

\end{abstract}

\maketitle

\section{Introduction}

\label{sec.einleitung3}

\begin{figure}
  \subfigure[]{%
    \includegraphics[width=\db]{\bilder/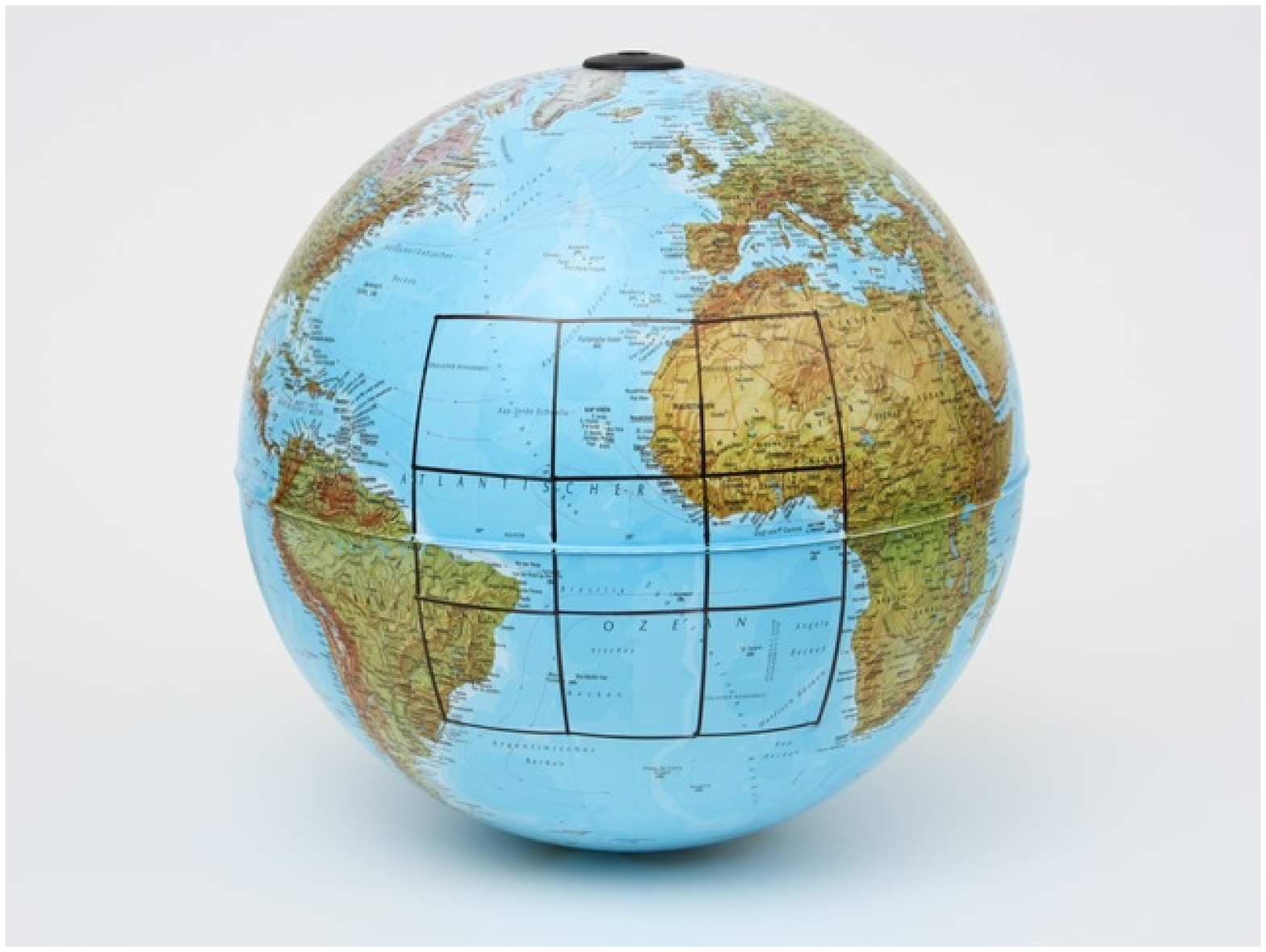}
  }\hfill%
  \subfigure[]{%
    \includegraphics[width=\db]{\bilder/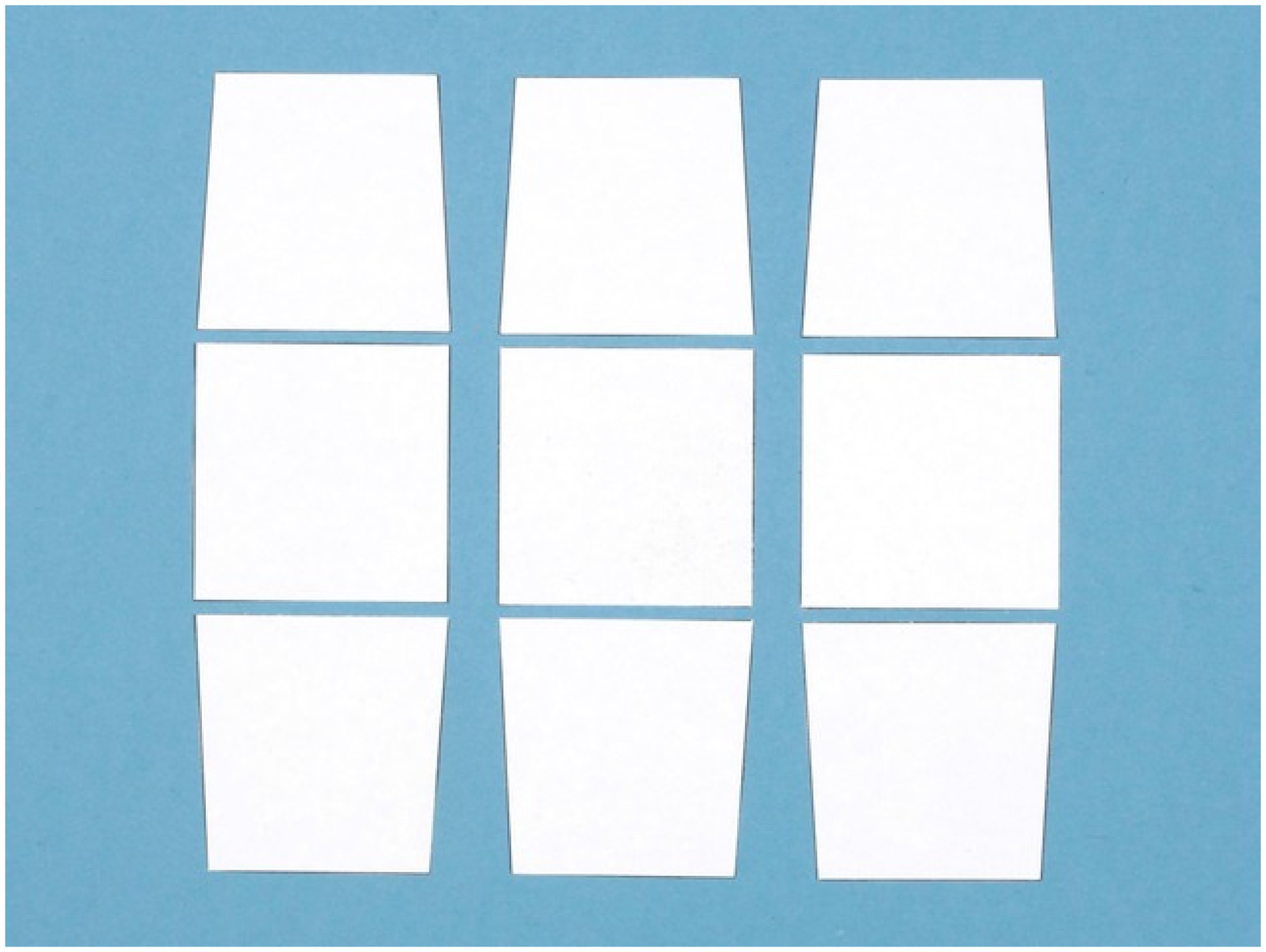}
  }\hfill%
  \subfigure[]{%
    \includegraphics[width=\db]{\bilder/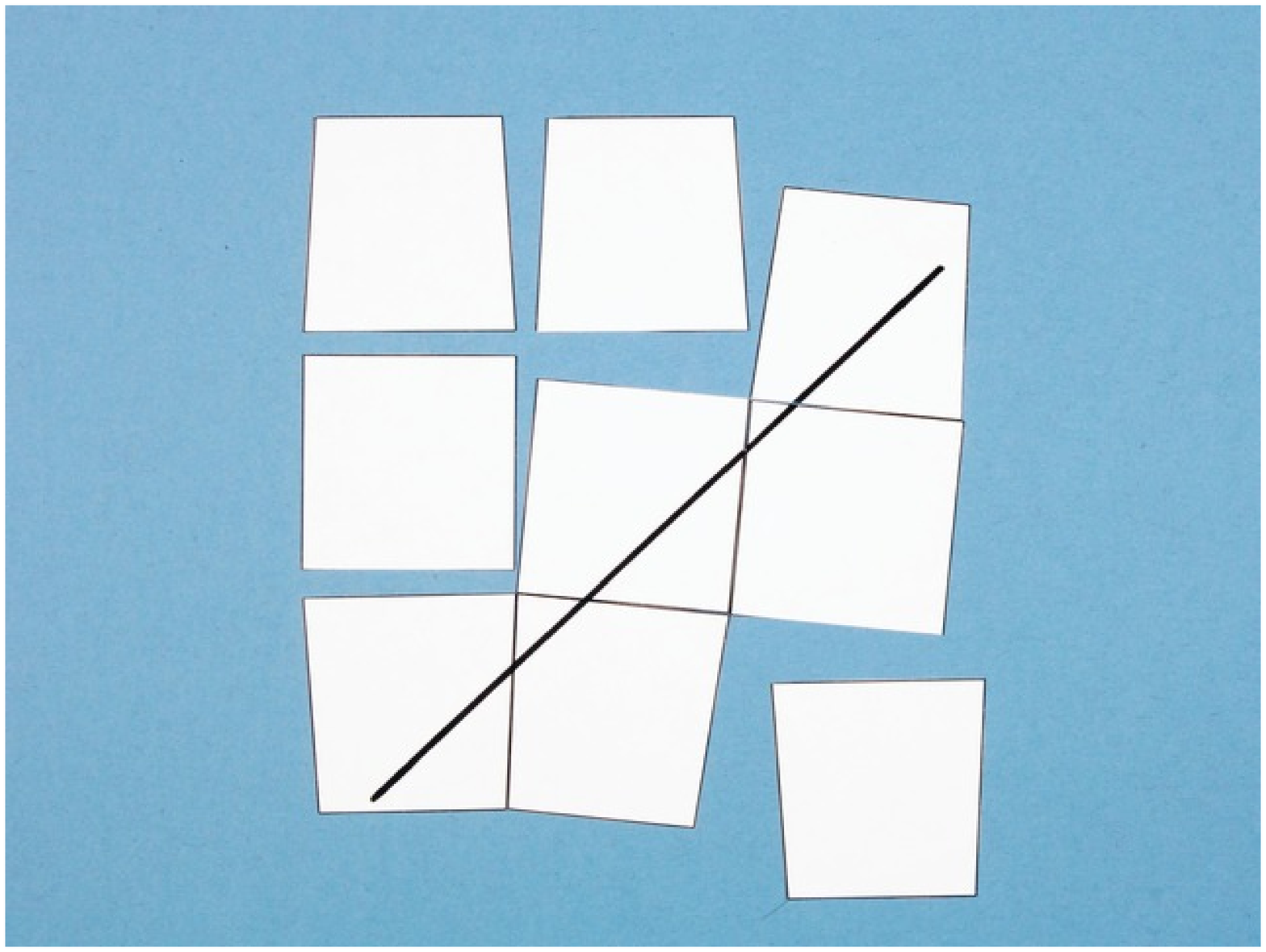}
  }
  \caption{\protect\raggedright
A sector model and the construction of a geodesic
using a sphere by way of example.
The curved surface is subdivided into small elements of area (a).
Their edge lengths are determined and the sectors are constructed
as flat pieces with the same edge lengths (b).
A geodesic is constructed as a locally straight line using pencil
and ruler (c).
  \label{fig.zusammen3}
  }
\end{figure}

In view of the goal
of teaching
introductory general relativity
without going beyond elementary mathematics,
we are developing an approach based on a particular class of
physical models, so-called sector models.
The approach relies on the fact
that general relativity is a geometric theory
and is therefore accessible to intuitive geometric understanding.
In the first part of this series,
we have developed sector models as physical models
of curved spaces and spacetimes  \citep[in the following referred to as paper~I]{teil1}.
Sector models implement the description of curved spacetimes
used in the Regge calculus \citep{reg1961}
by way of physical models.
Sector models
can be two-dimensional (e.g.\ a symmetry plane of a spherically
symmetric star),
three-dimensional (e.g.\ the three-dimensional curved space
in the exterior region of a black hole),
or 1+1-dimensional (i.e.\ a spacetime with two spatial dimensions
suppressed, similar to the Minkowski diagrams of special relativity).
Figure~\ref{fig.zusammen3} illustrates the basic principle
using a sphere by way of example:
The curved surface is subdivided into small elements of area,
in this case quadrilaterals (figure~\ref{fig.zusammen3}(a)).
The edge lengths are determined for each quadrilateral.
Quadrilaterals in the plane are constructed with the same edge lengths
(figure~\ref{fig.zusammen3}(b)).
These are the sectors that make up the sector model.
The sector model is an approximation to the curved surface,
its accuracy being determined by the resolution of the subdivision.
For pedagogical purposes, a relatively coarse resolution is useful.
Using sector models,
the geometry of the
respective
space or spacetime
can be studied with graphical methods.
This includes
the construction of geodesics
as described in the second paper of this series
\citep[in the following referred to as paper~II]{teil2}.
The construction implements the
determination of geodesics
according to the Regge calculus
\citep{wil1981}.
The basic principle is illustrated in figure~\ref{fig.zusammen3}(c).
Based on the definition of a geodesic as a locally straight line,
the geodesic is drawn using pencil and ruler:
Inside a sector, a sector being a flat element of area,
a geodesic is a straight line.
When the line reaches the border of the sector,
the neighbouring sector is joined
and the line is continued straight across the border.
Sector models are computed true to scale,
therefore, the properties read off from them
are quantitatively correct, within the discretization error.
The accuracy achievable for geodesics
is studied in paper~II.

General relativity describes the paths of light and free particles
as geodesics in spacetime.
This contribution
shows how geodesics in spacetime
can be constructed
using spacetime sector models as tools.
Radial geodesics close to a black hole
are used by way of example.
We outline two workshops
the way that we teach them for undergraduate students.
In the workshop on gravitational redshift
(section~\ref{sec.workshoprot}),
null geodesics are constructed
and the phenomenon of gravitational redshift
is inferred.
The workshop on radial free fall
(section~\ref{sec.workshopball})
includes the construction of radial timelike geodesics
and a comparison with the Newtonian descriptions of free fall
and of tidal forces.
Conclusions and outlook follow in section~\ref{sec.fazit3}.

\section{Workshop on gravitational redshift}
\label{sec.workshoprot}

In this workshop,
world lines of light are constructed
as geodesics in spacetime.
The construction shows how gravitational redshift arises.
A black hole is used by way of example,
because in its vicinity
the effects are large and are
clearly visible in the graphical representation.
The construction of geodesics is restricted to radial paths;
in the representation of the spacetime
the other two spatial directions are suppressed
so that the spacetime sector model is 1+1-dimensional.

The workshop presumes
that the participants are familiar with
the concept of a geodesic as a locally straight line
and with sector models as representations of surfaces with curvature.
A knowledge of special relativity
is not required
for this workshop.
Minkowski diagrams do play a role and
if necessary they
are explained
to the extent that they are needed in the workshop:
Firstly they are introduced as path-time diagrams
that differ from the familiar diagrams used in Newtonian mechanics
by the fact that the vertical axis is the time axis.
In order to familiarize the participants
with this representation,
we show a diagram with world lines
that tell a little story
and ask the participants to recount what happens
(an example of such a diagram is available online,~\citealp{teil3}).
Secondly, the units of the axes are addressed,
being chosen to make
motion at the speed of light
run at $45^{\circ}$ in the diagram.
Finally the terms event, world line and light cone are
introduced.

\subsection{Redshift close to a black hole}
\label{sec.rot}

\begin{figure}
  \subfigure[]{%
    \raisebox{0.5cm}{\includegraphics{\bilder/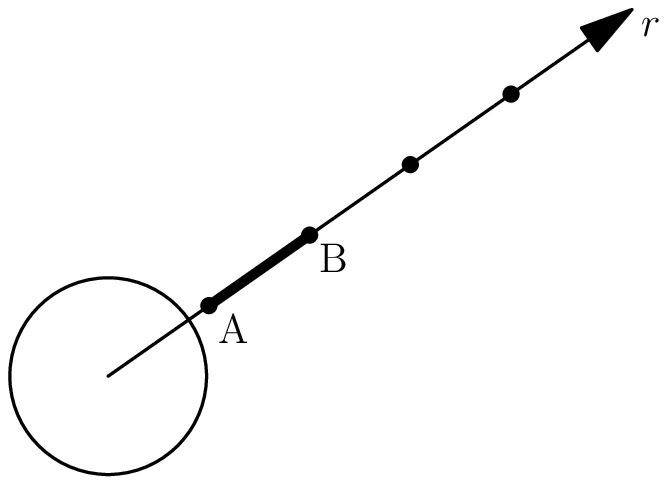}}
  }\hfill%
  \subfigure[]{%
    \includegraphics{\bilder/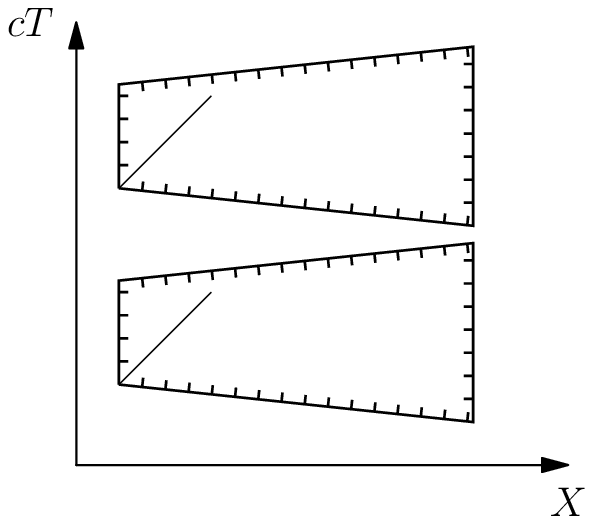}
  }
\caption{\label{fig.rtmodell}
Sector model for the spacetime of a radial ray
in the exterior region of a black hole.
(a) Radial ray.
(b) Spacetime sector model for the segment marked in (a).
The edge on the left
represents events at point~A,
the edge on the right
events at point~B.
The diagonal lines mark the light cone.
The model can be extended in time
by adding identical sectors.
}
\end{figure}
The workshop begins with the explanation
that general relativity
describes the paths of light and free particles
as geodesics in spacetime.
Then a sector model is introduced
that represents the spacetime of a radial ray
in the exterior region of a black hole.
The participants can calculate the sector model
themselves
(section~\ref{sec.trberechnung})
or can be provided with a worksheet
(available online, \citealp{teil3}).
In a thought experiment
the sector model is created
from measurements taken close to a black hole:
Astronauts travel to a black hole
and take up positions along a radial ray.
They choose a number of events at these positions
and use them as vertices
for subdividing the spacetime
of the radial ray into quadrilaterals.
In order to define an individual quadrilateral,
two positions are chosen on the radial ray
(figure~\ref{fig.rtmodell}(a)).
Two events at the inner position
and two at the outer position
make up the four vertices.
Each quadrilateral
is represented by a sector of a Minkowski space
(figure~\ref{fig.rtmodell}(b));
the ensemble of sectors makes up the sector model.%
\footnote{The sector model covers the region
from $1.25$ to $2.5$ Schwarzschild radii in the Schwarzschild
radial coordinate, see section~\ref{sec.trberechnung}.}
Since the black hole spacetime is static
(we consider a non-rotating black hole),
it is possible to choose a subdivision of the spacetime
for which the shape of the sectors is independent of time
(see section~\ref{sec.trberechnung}).
This is here implemented,
so that the sector model can be arbitrarily extended
in time by adding identical sectors.

\begin{figure}
\begin{center}
\includegraphics{\bilder/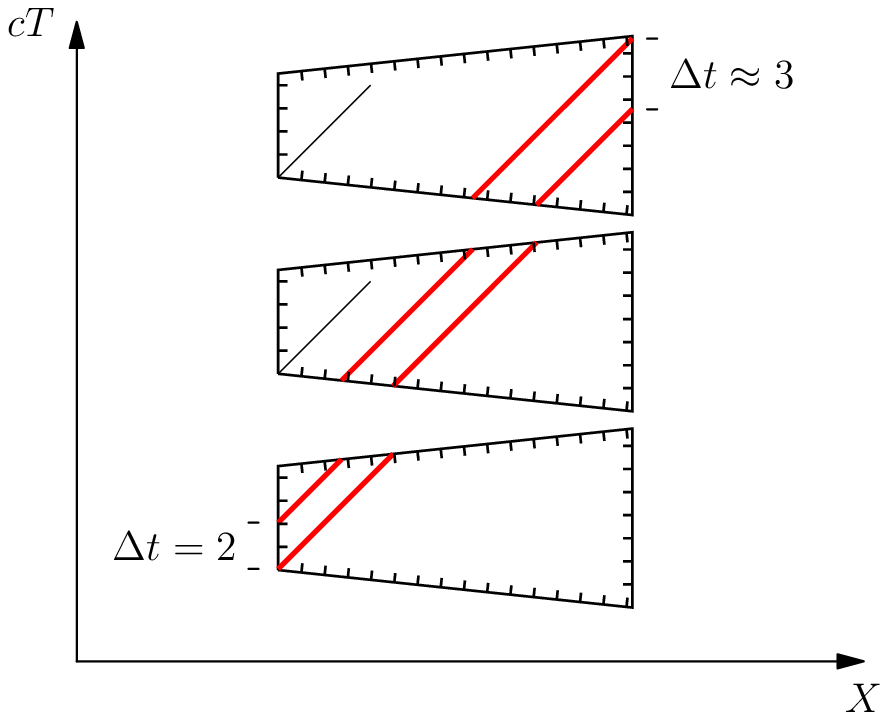}
\end{center}
\caption{\label{fig.rot}
World lines of two light signals
propagating radially outwards.
An observer at the inner rim of the sectors (left)
sends the signals at an interval of two time units.
An observer at the outer rim (right) receives them
about three time units apart.
}
\end{figure}

The first construction on the sector model
is
the world line of a light signal
propagating radially outwards.
Starting from
the bottom left corner of the model,
a locally straight line is drawn
in the direction of the light cone
(figure~\ref{fig.rot}, bottom line).
Within the sector, the line is straight.
When the line reaches the border of the sector,
the position of the border point is
copied onto the respective border of the neighbouring
sector and the line is continued from there.
The borders are provided with
equidistant tick marks
to facilitate the transfer of the border points.
The direction in the neighbouring sector
is again prescribed by the light cone,
since we are concerned with a world line of light.\footnote{%
Alternatively, one may join the neighbouring sector
and then continue the line straight across the border
as described in
figure~\protect\ref{fig.zusammen3}.
How to join spacetime sectors is described in
section~\ref{sec.zeitartig}.}

In the second step
the transmission of two consecutive
light signals is studied.
An observer positioned close to the black hole
and at a constant distance
(point~A at the inner rim of the
sectors, see figure~\ref{fig.rtmodell}),
sends two light signals outwards,
one a short time after the other.
In figure~\ref{fig.rot} the time interval
is two time units.
A second observer
positioned farther away from the black hole
and also at a constant distance
(point~B at the outer rim of the sectors),
receives the two signals.
In order to find the time interval of the light signals
upon reception,
the world line of the other signal is added to the diagram
(figure~\ref{fig.rot}, top line).
The interval of the signals can then be read off
and amounts to a little over three time units.
If one interprets the two signals
as consecutive wave crests
of an electromagnetic wave,
one concludes from the diagram
that the wave is received
with an increased period
by the outer observer.
Thus, radiation receding from a black hole
is redshifted.
The ratio of the periods $P_{\rm outer}$ and $P_{\rm inner}$
at points~B and~A, respectively,
read off from the
construction on the sector model, is
$P_{\rm outer} / P_{\rm inner} \approx 1.5$.
The calculated exact value is
$P_{\rm outer} / P_{\rm inner}
   = \sqrt{(1-\rs/r_{\rm outer})/(1-\rs/r_{\rm inner})}
   = 1.73$,
where
$r_{\rm inner} = 1.25\,\rs$
and
$r_{\rm outer} = 2.5\,\rs$
are the radial coordinates of points~A and~B, respectively.
The graphically determined value is too small by 13\%;
this deviation is due
to the relatively coarse resolution of the sector model.

\subsection{Calculation of the spacetime sector model}
\label{sec.trberechnung}

A simplified calculation of sector models
is introduced in paper~II (section~2.4)
for curved surfaces
and is here extended
to the 1+1-dimensional case.
This calculation presumes knowledge of the Minkowski metric.
Using the simplified method
students can calculate sector models
on their own
using elementary mathematics only.
This enables them to
use sector models as tools for studying
other curved spacetimes when given their metric.
The approximations of the simplified method
are discussed in paper~II.

The starting point of the construction is the metric
\begin{equation}
\label{eq.ssm-rt}
\ud s^2 = - \left (1 - \frac{\rs}{r} \right) c^2 \ud t^2 +
            \frac{1}{1 - \rs/r } \ud r^2
\end{equation}
with the usual Schwarzschild coordinates $t$ and $r$
and the Schwarzschild radius  $\rs = 2 G M / c^2$
of the central mass $M$, where $G$ is the gravitational constant
and $c$ the speed of light.

\begin{figure}
  \centering
  \subfigure[]{%
    \includegraphics{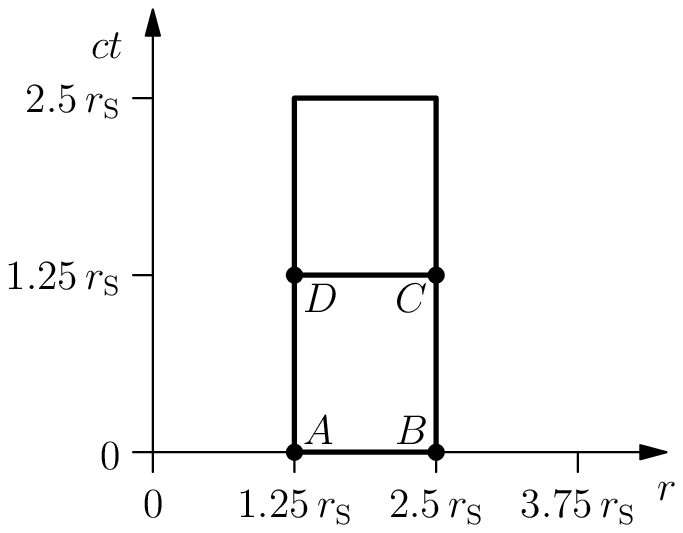}
  }\hfil%
  \subfigure[]{%
    \includegraphics{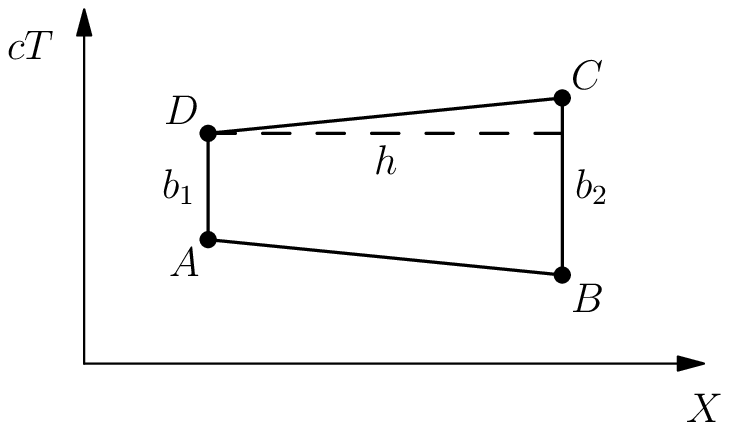}
  }
  \caption{\label{fig.rtteilung}
Calculation of the spacetime sector model of a radial ray.
(a) The subdivision of the spacetime in coordinate space.
(b) Each sector is constructed in the shape of a symmetric trapezium.
}
\end{figure}

In space the sector model represents a segment of a radial ray
from $r=1.25\, \rs$ to $r=2.5\, \rs$
with coordinate length $\Delta r = 1.25\, \rs$.
The time coordinate $t$ is subdivided into segments of length
$\Delta t$ with
$c\Delta t = 1.25\, \rs$ (figure~\ref{fig.rtteilung}(a)).
Since the metric is independent of the time coordinate,
only a single segment in time needs to be calculated.

The calculation of the edge intervals
yields
\begin{equation}
  \Delta s_t^2 (r) = - \left (1 - \frac{\rs}{r} \right) c^2 \Delta t^2
     \hspace*{1cm} (\Delta r = 0)
\end{equation}
for the timelike edges and
\begin{equation}
\Delta s_r^2 = \frac{1}{(1-\rs/\rrmi)}\, \Delta r^2
     \hspace*{1cm} (\Delta t = 0)
\end{equation}
for the spacelike edges,
where
the metric coefficient
is evaluated at the mean coordinate $\rrmi=(r_1+r_2)/2$
with the coordinates $r_1$ and $r_2$ of the associated vertices.
Since
the sector model is a column of identical
sectors,
each sector is constructed with the corresponding
time symmetry:
As shown in figure~\ref{fig.rtteilung}(b),
the bases of a trapezium
are drawn in direction of the time axis
with the lengths
$b_1=\sqrt{-\Delta s_t^2 (r=1.25\,\rs)}$
and $b_2=\sqrt{-\Delta s_t^2 (r=2.5\,\rs)}$.
The height $h$ of the trapezium
is determined from the condition that
the lateral sides have the interval
$\Delta s_r^2$ (figure~\ref{fig.rtteilung}(b)):
\begin{equation}
\Delta s_r^2
= - \left ( \frac{b_2 - b_1}{2} \right )^2 + h^2.
\end{equation}
The result is the sector shown in figure~\ref{fig.rtmodell}.

\section{Workshop on particle paths}
\label{sec.workshopball}

In this workshop
world lines are constructed
for freely falling particles
close to a black hole.
As in the previous section
the study is restricted to radial paths,
so that it is possible to use a
1+1-dimensional spacetime sector model.
By means of the particle paths,
the connection between the relativistic and the
classical descriptions
of motion in a gravitational field is pointed out.
The workshop presumes that the participants
are familiar with the Lorentz transformation.

\begin{figure}
  \centering
  \subfigure[]{%
    \includegraphics{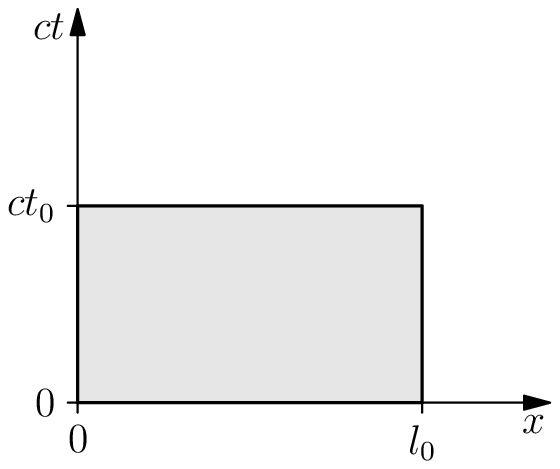}
  }\hfil%
  \subfigure[]{%
    \includegraphics{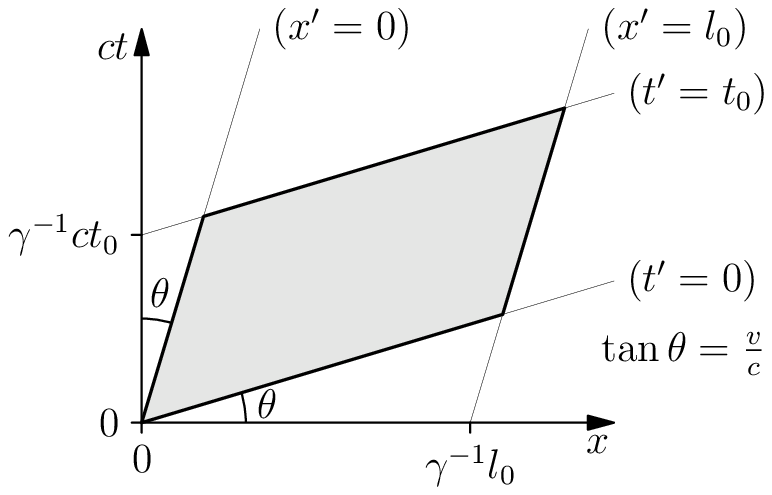}
  }
\caption{\label{fig.raumschiff}
Graphic representation of a spacetime sector
in two different reference frames.
The events are located inside a spaceship of length $l_0$
at spaceship proper times between zero and $t_0$.
(a) Representation in the rest frame of the spaceship.
(b) Representation in the rest frame of a space station
   that the spaceship passes with velocity $v=0.3\,c$
($\gamma=1 / \sqrt{1-v^2/c^2}$).
}
\end{figure}

\subsection{The construction of timelike geodesics}
\label{sec.zeitartig}

To study the paths of freely falling particles
in the vicinity of a black hole,
their world lines are constructed on a sector model.
As in the previous examples,
the geodesics are drawn as straight lines within each sector
and after reaching the boundary
are continued in the neighbouring sector.
Other than for the null geodesics
discussed in section~\ref{sec.workshoprot},
the direction in the neighbouring sector
is not predetermined by the light cone.
Therefore, it is necessary to join the neighbouring sector
and to continue the line straight across the boundary.
The joining of two sectors is more complex in spacetime
than in the purely spatial case.
Clearly, rotating the neighbouring sector
into a suitable position is not an option:
Since the speed of light has the same value in both sectors,
their light cones must coincide. This fixes the orientation
of the neighbouring sector.

In the workshop we use a specific example for a spacetime sector
in order to introduce the way that
neighbouring sectors can be
joined.
We consider a long and very thin spaceship of rest length $l_0$.
We define a spacetime sector that is made up of all events
that are inside the spaceship
and at spaceship proper times between zero
and $t_0$.
The participants draw this spacetime sector in a Minkowski diagram,
first in the rest frame of the spaceship (figure~\ref{fig.raumschiff}(a)),
then in the rest frame of a space station that the spaceship passes
at constant relative velocity 
(figure~\ref{fig.raumschiff}(b)).
The geometric shape of the sector,
i.e., the geometric shape drawn on paper
and understood in the euclidean sense,
clearly depends on the frame of reference.
Figures~\ref{fig.raumschiff}(a) and~(b) are two different representations
of one and the same spacetime sector.
One turns into the other under a Lorentz transformation.

\begin{figure}
  \centering
  \includegraphics[scale=0.5]{\bilder/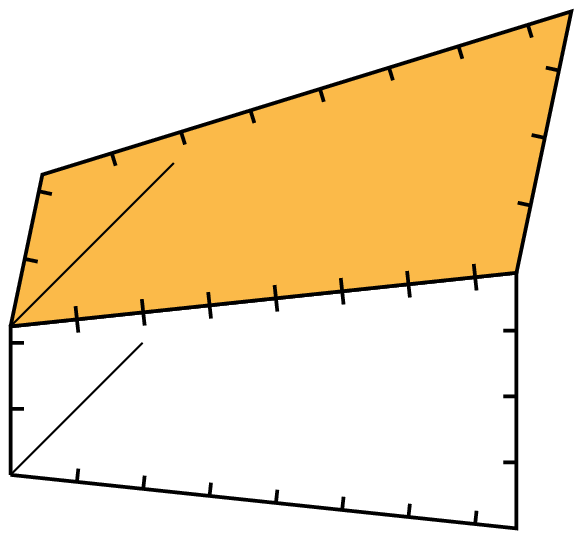}
  \caption{\label{fig.trtransfersektor}
Joining two spacetime sectors of the sector model
shown in figure~\protect\ref{fig.rtmodell}.
The upper sector
has been Lorentz-transformed in such a way
that it can be joined to the lower sector
($v/c = 0.21$).
}
\end{figure}

\begin{figure}
  \centering
  \subfigure[]{%
    \includegraphics[width=\db]{\bilder/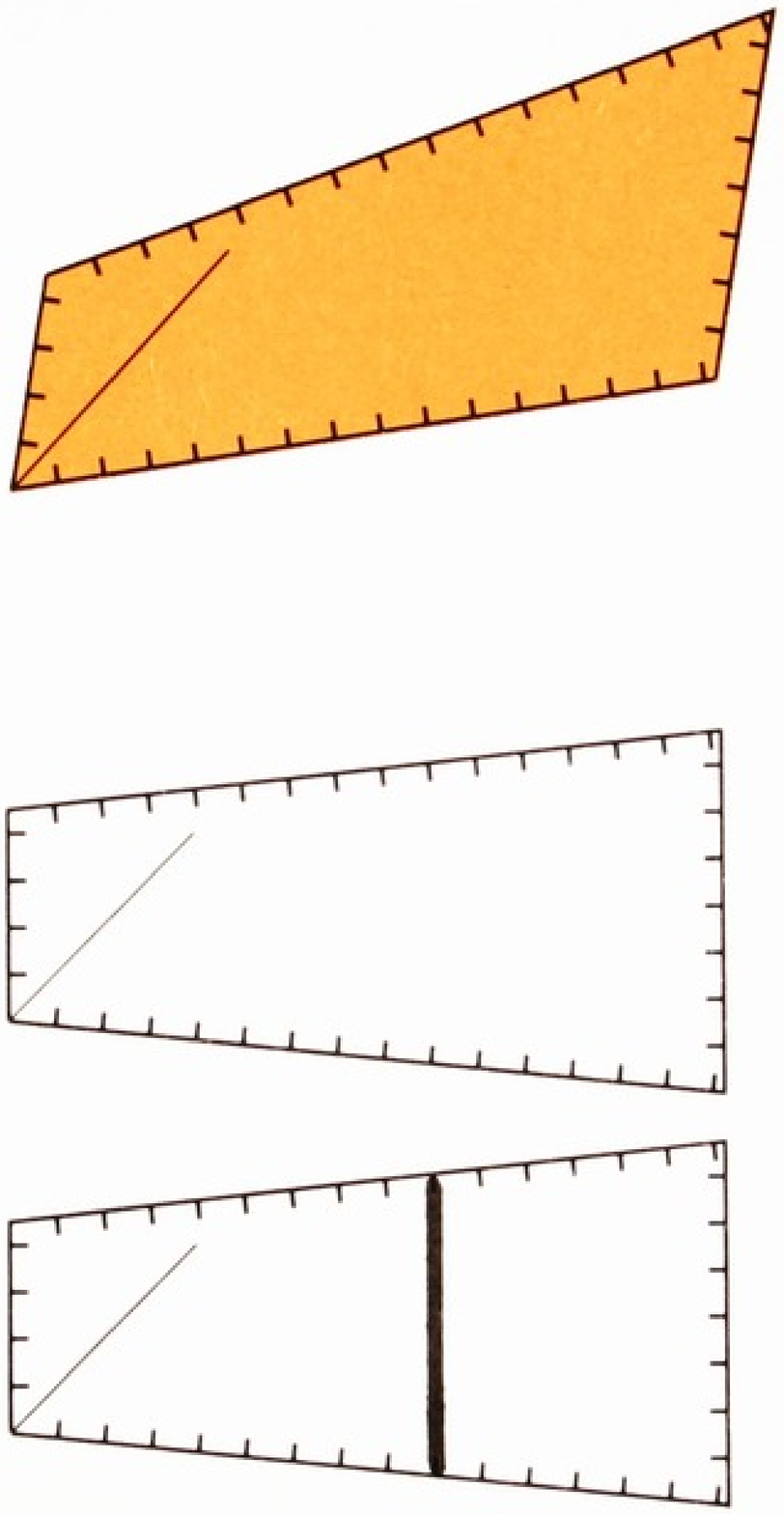}
  }\hfill%
  \subfigure[]{%
    \includegraphics[width=\db]{\bilder/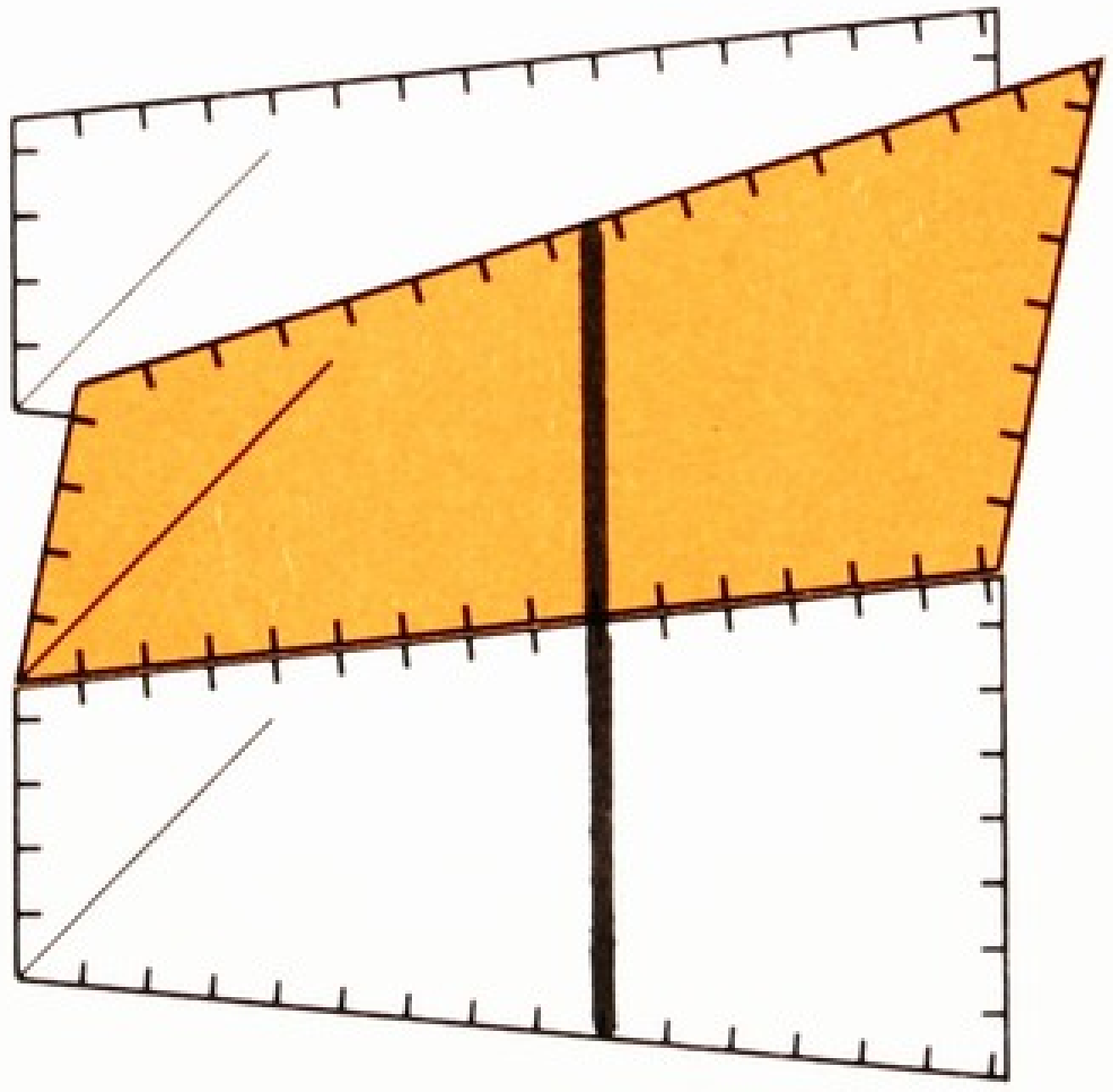}
  }\hfill%
  \subfigure[]{%
    \includegraphics[width=\db]{\bilder/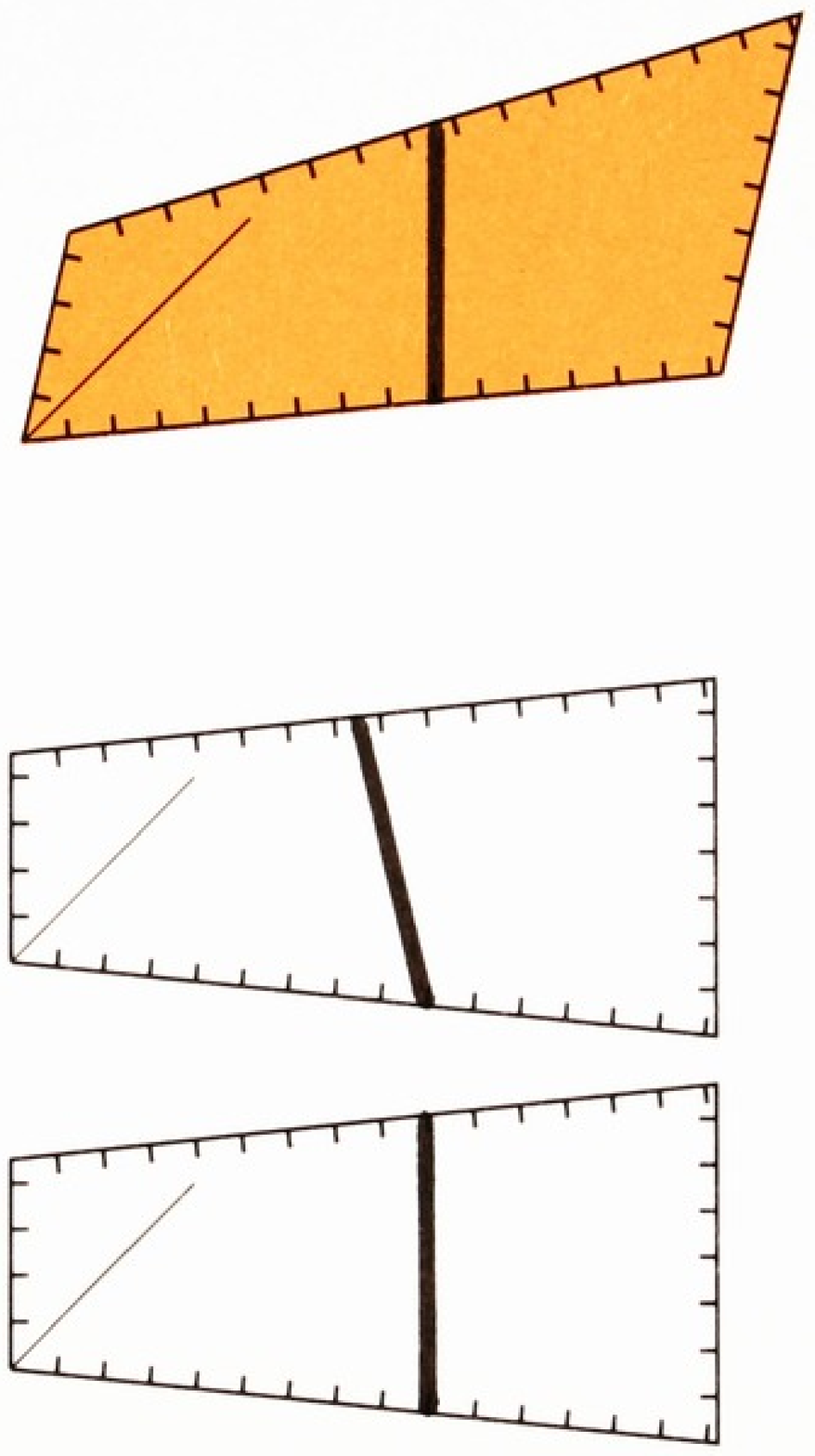}
  }
  \caption{\label{fig.trgeodzeichnen}
Construction of a geodesic on the spacetime sector model.
(a) A sector that has undergone Lorentz transformation
serves as transfer sector (in colour).
(b) The geodesic is continued straight onto the transfer sector.
(c) The line is copied from the transfer sector onto the neighbouring
sector in the original symmetric shape.
}
\end{figure}

This permits the joining of sectors:
The neighbouring sector is brought into the appropriate form
by Lorentz transformation
(figure~\ref{fig.trtransfersektor}).
Thus, the transformation that permits the joining of the neighbouring sector
is a rotation in the spatial case and a Lorentz transformation
in the spacetime case.

When drawing geodesics,
it is convenient to use a transformed sector
in the role of transfer sector (paper~II, section 2.3):
When a geodesic reaches the border of a sector
(figure~\ref{fig.trgeodzeichnen}(a)),
it is continued as a straight line across the border
onto the transfer sector joined to the respective edge
(figure~\ref{fig.trgeodzeichnen}(b))
and is then transferred
onto the neighbouring sector
in its original shape
(figure~\ref{fig.trgeodzeichnen}(c)).
This transfer amounts to reversing the Lorentz transformation.
In doing so, straight lines are mapped onto straight lines.
Therefore, using the tick marks at the borders,
the end points of the line
are transferred onto the target sector and are then
connected by a straight line (figure~\ref{fig.trgeodzeichnen}(c)).

\begin{figure}
  \centering
  \subfigure[]{%
    \includegraphics[scale=0.4]{\bilder/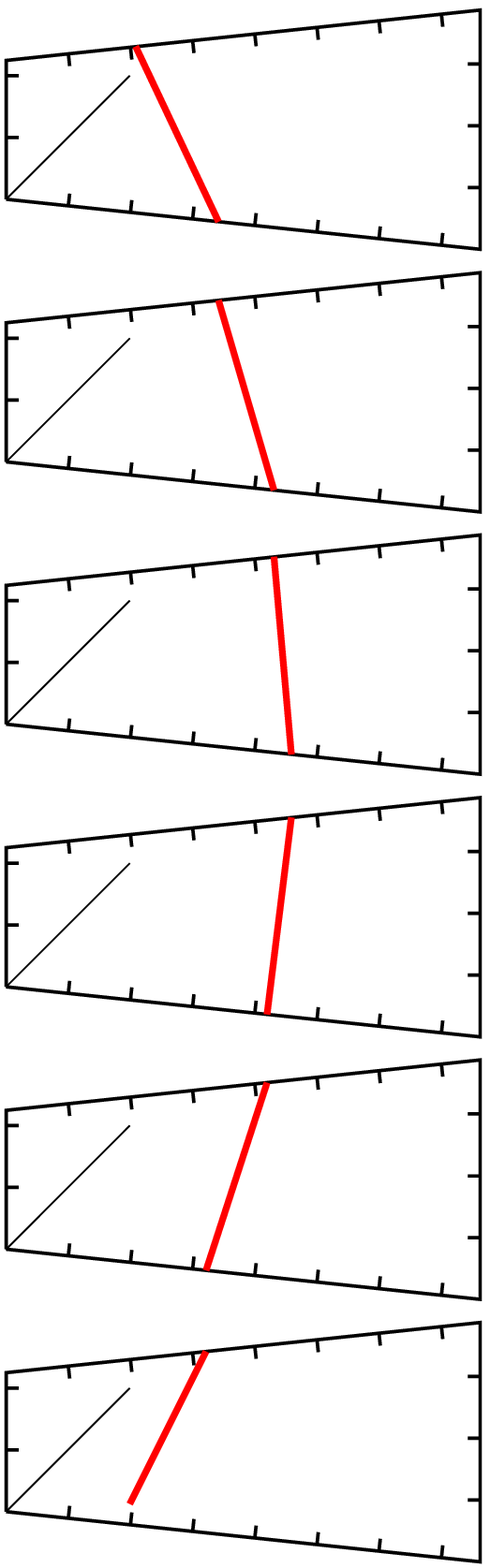}
  }\hspace*{2cm}%
  \subfigure[]{%
    \includegraphics[scale=0.4]{\bilder/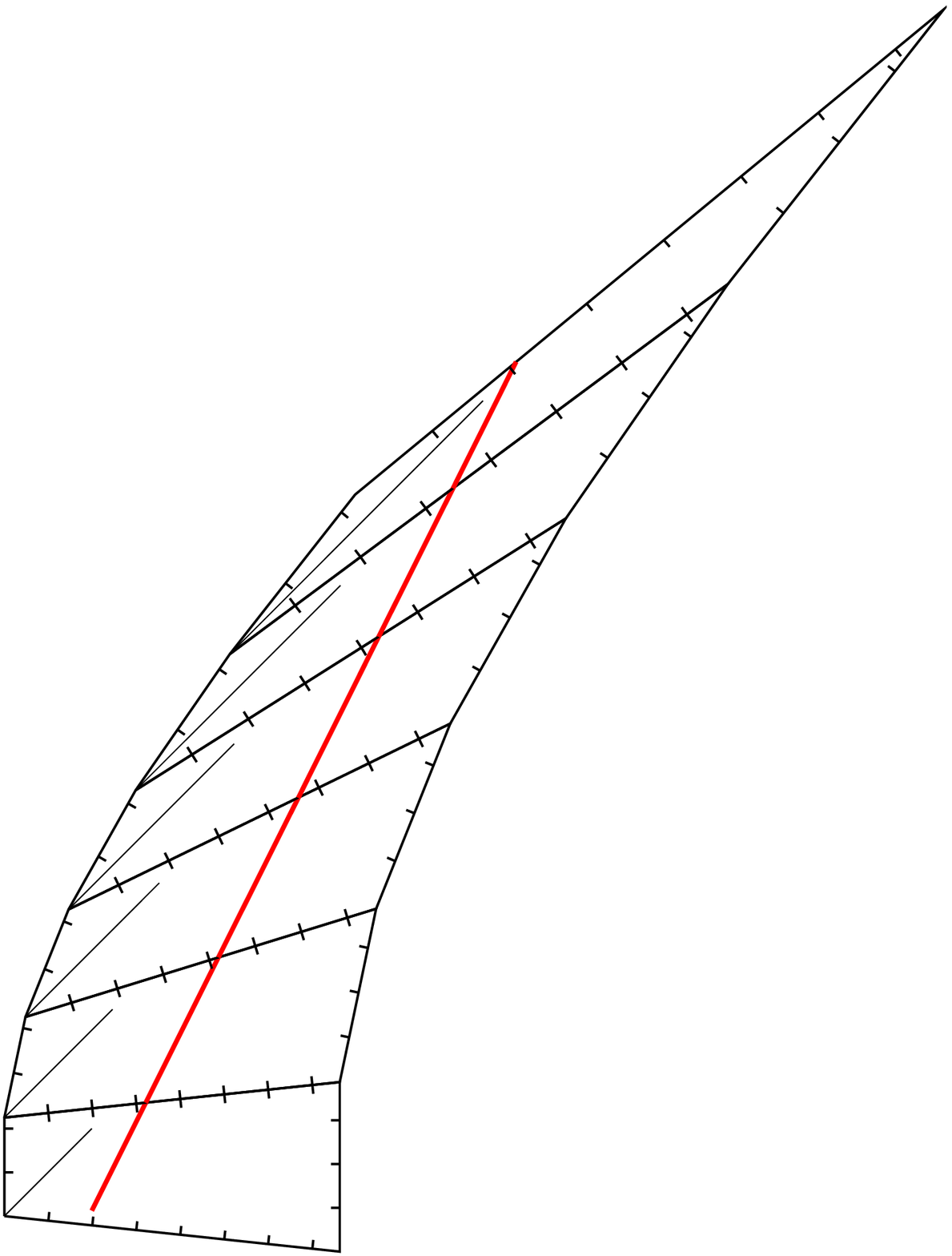}
  }
  \caption{\label{fig.wurf}
Vertical free fall.
(a) A geodesic constructed on the sector model.
(b) When all sectors
are joined,
the world line can clearly be seen to be straight.
}
\end{figure}

\subsection{Vertical free fall}
\label{sec.wurf}

Close to a black hole, a particle is thrown upwards.
Its path is to be determined.
Intuitively, it is clear
that the particle reaches a maximum
height and then falls back down
(provided its initial velocity is less than the
escape velocity).

In the relativistic description,
the particle, being in free fall, follows a geodesic,
i.e., its world line is locally straight.
How can these two statements---straight world line on the one hand
and up-and-down motion on the other hand---be compatible?

For the construction of the world line,
the sector model shown in figure~\ref{fig.rtmodell}
is used with six rows
plus an appropriately transformed transfer sector
(figure~\ref{fig.trtransfersektor}).
After choosing
an initial position and a timelike outward direction,
the world line is constructed as a geodesic on the sector model
(figure~\ref{fig.wurf}(a)):
The locally straight line at first
leads away from the black hole and
then comes closer again.
The spacetime geodesic thus provides the expected
up-and-down motion in space.
In addition,
figure~\ref{fig.wurf}(b) shows the sector model
with all the sectors
joined,
so that the straightness of the line is obvious.
In order to draw the geodesic at a stretch
as in this figure,
one needs several representations of the sector
that are obtained by Lorentz transformations
with different velocities.
The construction on the sector model displays both the
straight line in spacetime and the up-and-down motion
in space, and so makes the connection
between them quite clear.

\begin{figure}
  \centering
  \subfigure[]{%
    \includegraphics{\bilder/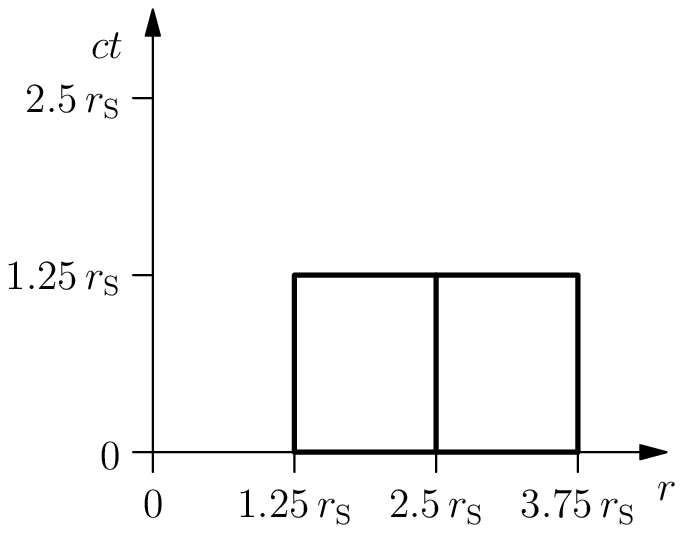}
  }\qquad
  \subfigure[]{%
    \includegraphics[scale=0.5]{\bilder/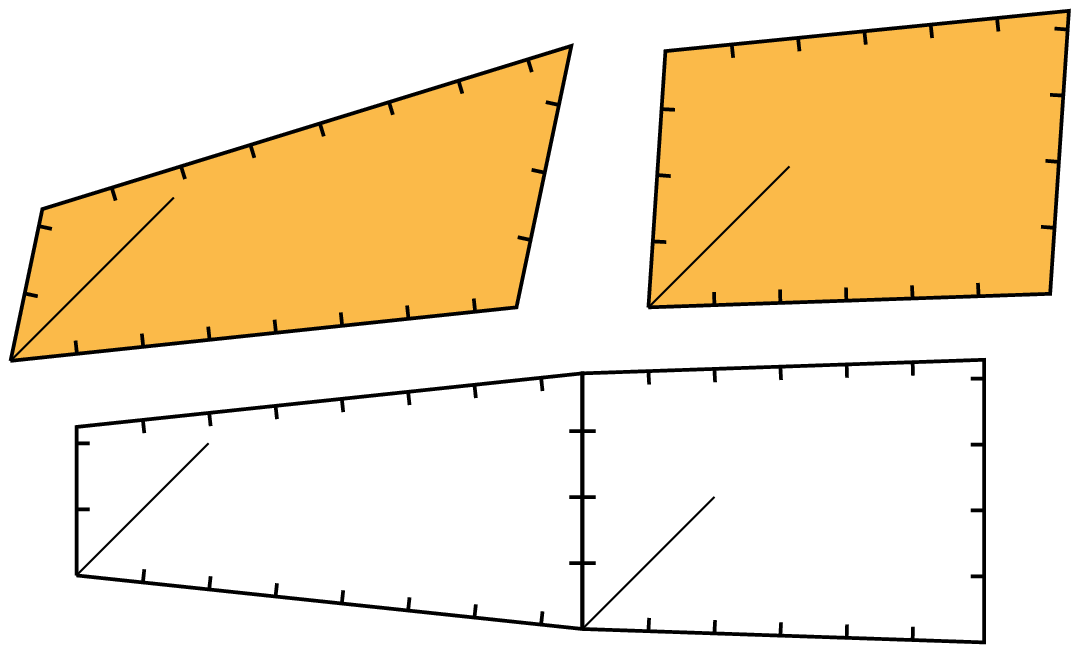}
  }
  \caption{\label{fig.rtmodell2}
Spacetime sector model for
a radial ray in the exterior region of a black hole.
(a) Subdivision of the spacetime in coordinate space.
(b) Sectors in symmetric form (bottom) and
suitably Lorentz-transformed transfer sectors (top).
This is an extension of the model shown
in figure~\protect\ref{fig.rtmodell}
by a second column with the associated transfer sector
(right hand side, $v/c = 0.067$).
The model can be extended in time
by adding identical rows.
}
\end{figure}

\subsection{Tidal forces and the curvature of spacetime}
\label{sec.gezeiten}

When the present workshop on particle paths is combined
with the workshop on curvature described in paper~I,
it is possible to illustrate the physical significance
of spacetime curvature with the help of geodesics.
For this purpose
a second column is added to
the sector model shown in
figure~\ref{fig.rtmodell},
so that the model now covers the radial ray
between $1.25\, \rs$ and $3.75\, \rs$ in two columns
(figure~\ref{fig.rtmodell2}).
The model is used with eight rows
plus a transfer sector for each column.

In a local inertial frame momentarily at rest
with respect to the black hole,
we consider two particles
that are slightly displaced in the radial direction.
They are released from rest simultaneously,
so that they fall one after another
radially into the black hole.
In the classical description
the gravitational force decreases outwards.
Therefore, at each instant
the
outer particle experiences
a smaller acceleration than the
inner
one,
so that the two freely falling particles
are accelerated relative to each other:
As a result of tidal forces,
the relative velocity of
the particles increases.

\begin{figure}
\centering
  \includegraphics[scale=0.35]{\bilder/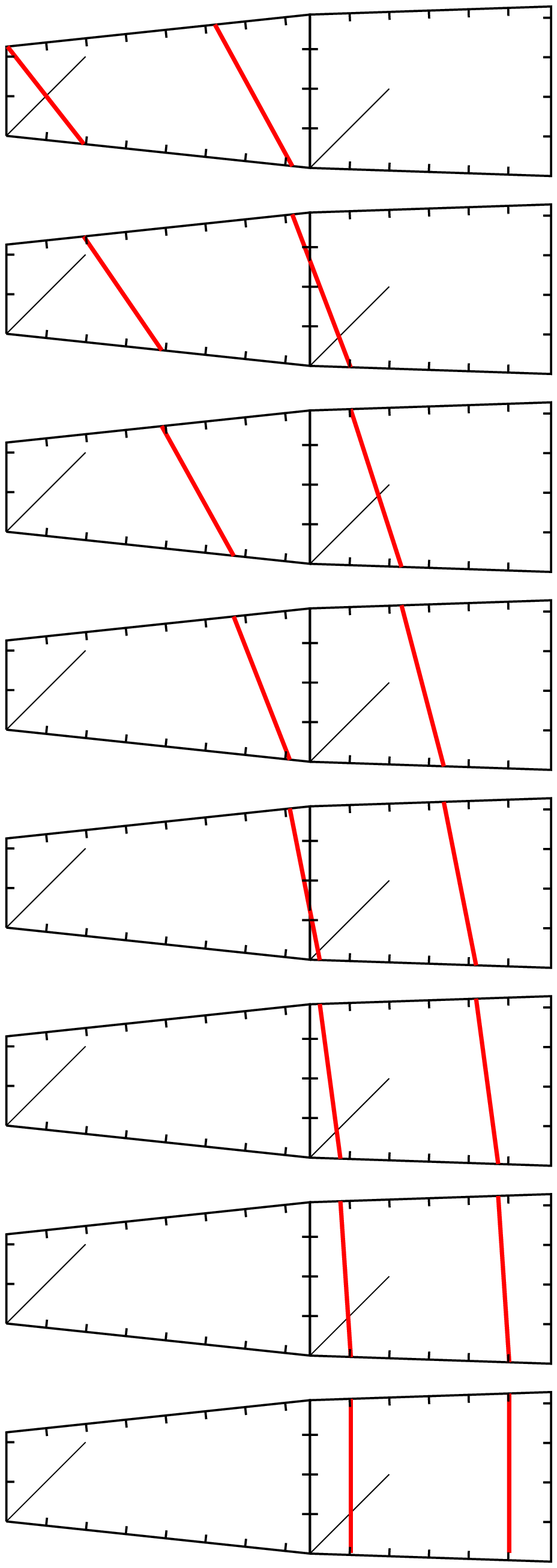}
  \caption{\label{fig.gezeiten}
The world lines of two particles
that are simultaneously released from rest and
fall one after another
towards a black hole.
}
\end{figure}

In the general relativistic description,
the world lines of the two particles are geodesics
that are initially parallel.
These geodesics are constructed on the sector model
(figure~\ref{fig.gezeiten}).
Both world lines start
in the direction of the local time axis
(figure~\ref{fig.gezeiten}, bottom row).
The two initially parallel world lines diverge more and more
indicating a relative velocity that increases.

The construction elucidates
the origin of the divergence:
The world lines are parallel
up to the point
where they pass a vertex on different sides
(figure~\ref{fig.gezeiten}, row 4 to row 5 from the bottom).
Each additional vertex
between the world lines
increases the difference in direction,
i.e., increases the relative velocity.

Figure~\ref{fig.vertexgezeiten}
shows the run of a pair of geodesics
close to a single vertex
in more detail.
For clarity, sectors
with double coordinate length
in time are used ($c\Delta t = 2.5\,\rs$).
In figures~\ref{fig.vertexgezeiten}(a) and~(b)
the sectors are joined along the geodesic on the left
and on the right, respectively
(the upper row
being suitably Lorentz-transformed
as a whole in each case);
in figure~\ref{fig.vertexgezeiten}(c)
the sectors are arranged symmetrically.

As described in paper~I, in a sector model
the so-called deficit angles of the vertices
represent curvature.
The deficit angle of the vertex considered here
is apparent in figures~\ref{fig.vertexgezeiten}(a) and~(b)
as the gap that remains when the four adjacent sectors
are joined around the vertex.
This deficit angle is in a spacelike direction and is
positive\footnote{%
The deficit angle is positive if
a wedge-shaped gap remains after joining
all adjacent sectors.
It is negative
if, after joining all the sectors except one,
the remaining space is too small to accommodate the last sector.};
with the metric signature used here,
by convention,
this means positive  spacetime curvature.
By construction,
the angle between the two lines behind the vertex
depends on the deficit angle.
Thus, figures~\ref{fig.gezeiten} and~\ref{fig.vertexgezeiten} show
that positive spacetime curvature
is linked with the divergence
of initially parallel world lines;
the opposite holds in the case of negative curvature.
Spacetime curvature is less intuitive than spatial curvature.
However, the
run of neighbouring geodesics
provides a criterion
that can be understood geometrically.
Thus the construction shows
how the relative acceleration
of the two particles
comes about
in the relativistic description.
It can be traced back to the deficit angles at the vertices,
i.e.\ to curvature.
This elucidates the physical meaning of spacetime curvature:
It corresponds to the Newtonian tidal force.

In addition,
figure~\ref{fig.vertexgezeiten}(d)
shows the behaviour of initially parallel spacelike geodesics
near the same vertex: After the vertex they converge.
The opposing behaviour
of timelike and spacelike geodesics
reflects the corresponding properties of
the deficit angles in timelike and spacelike directions,
respectively
(paper~I, section~4).
\newcommand{\figelfscale}{0.30}

\begin{figure}
  \centering
  \subfigure[]{%
    \includegraphics[scale=\figelfscale]{\bilder/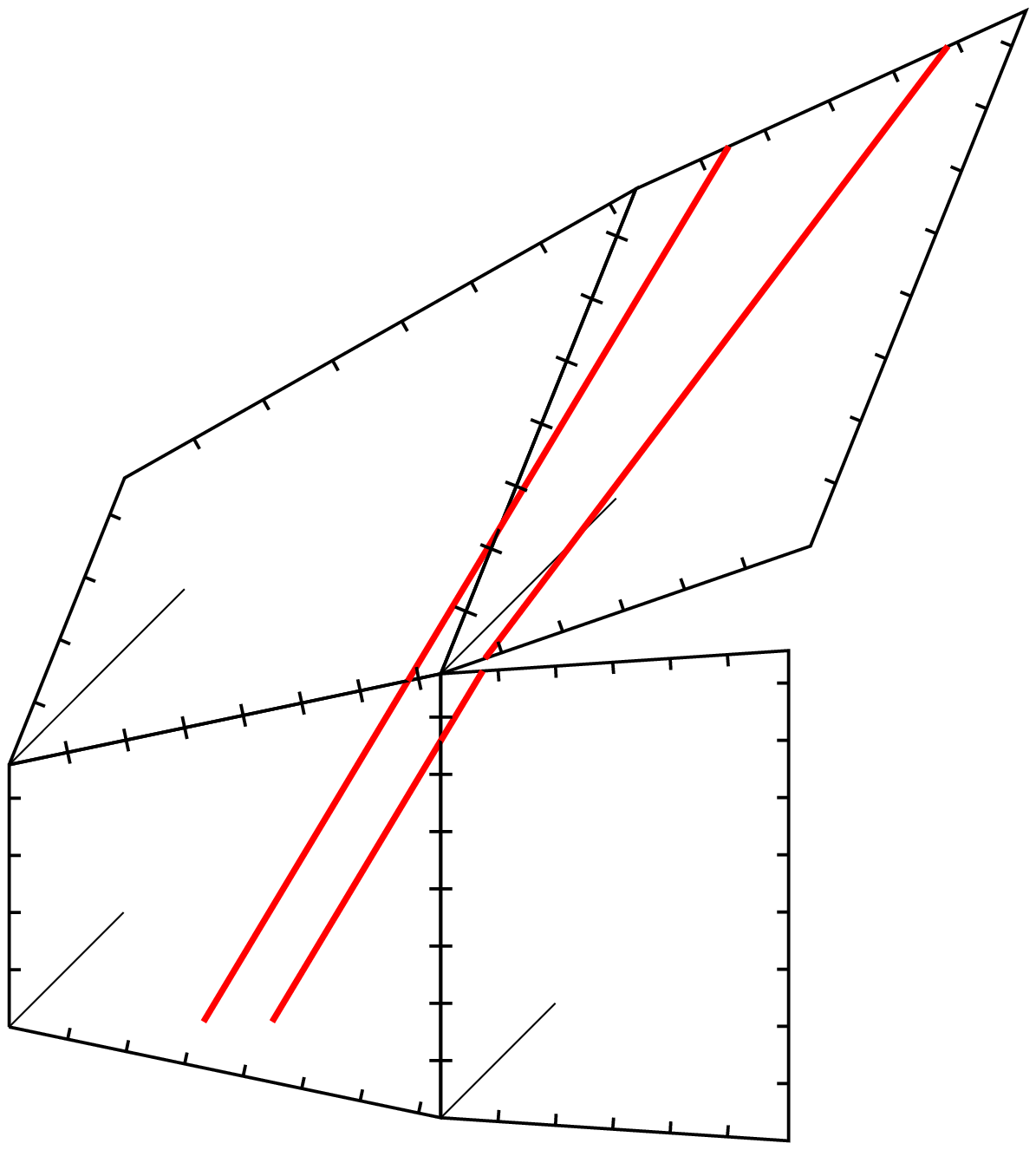}
  }\hfill%
  \subfigure[]{%
    \includegraphics[scale=\figelfscale]{\bilder/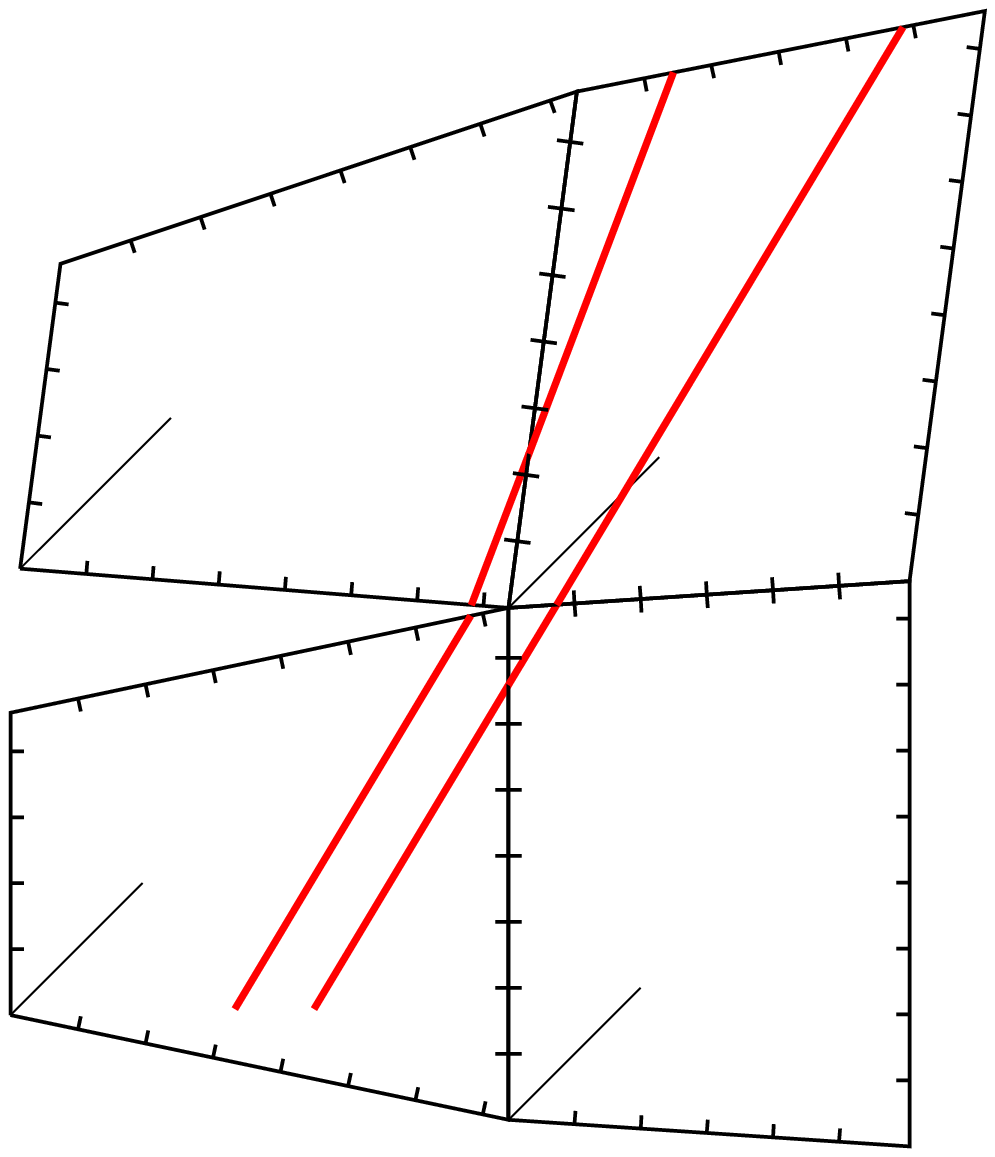}
  }\hfill%
  \subfigure[]{%
    \includegraphics[scale=\figelfscale]{\bilder/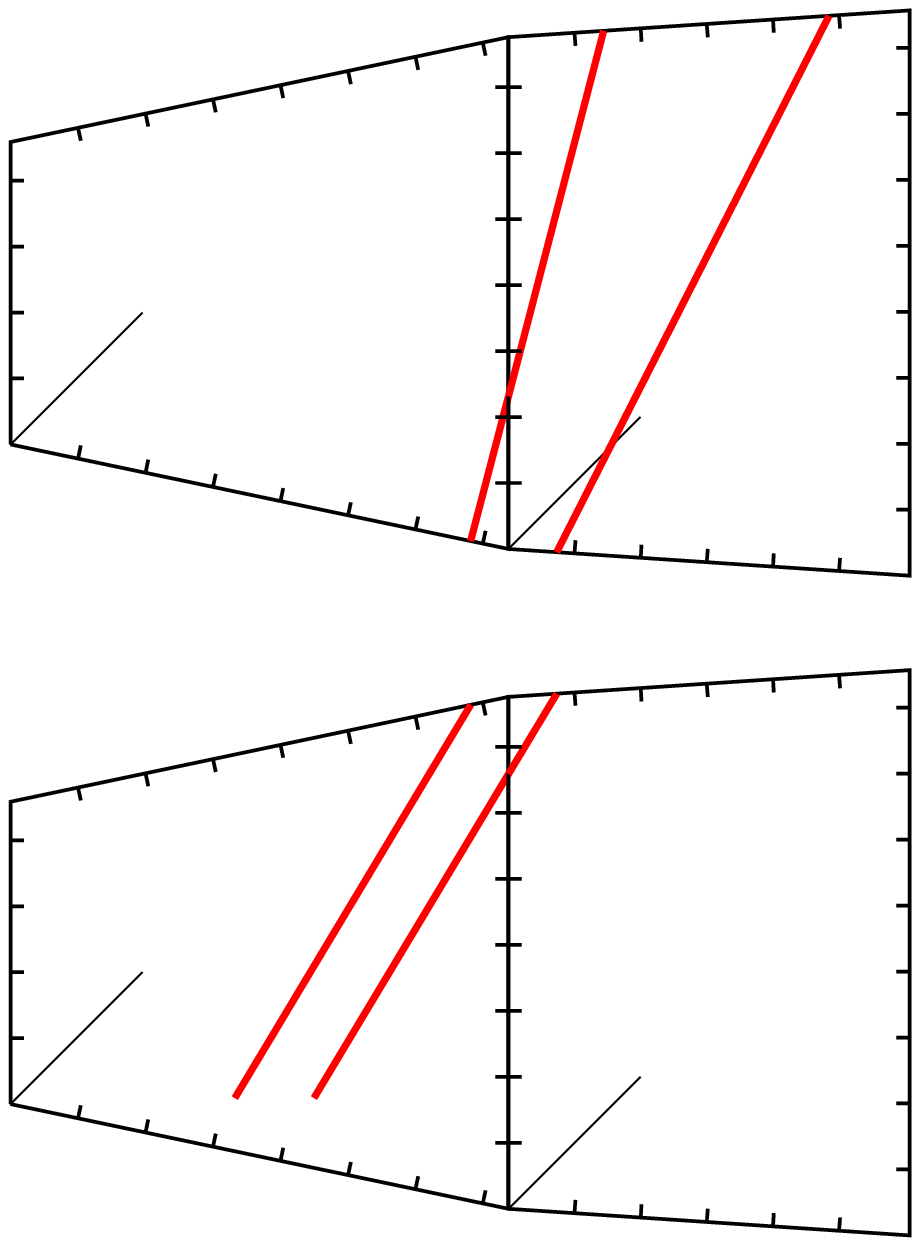}
  }\hfill%
  \subfigure[]{%
    \includegraphics[scale=\figelfscale]{\bilder/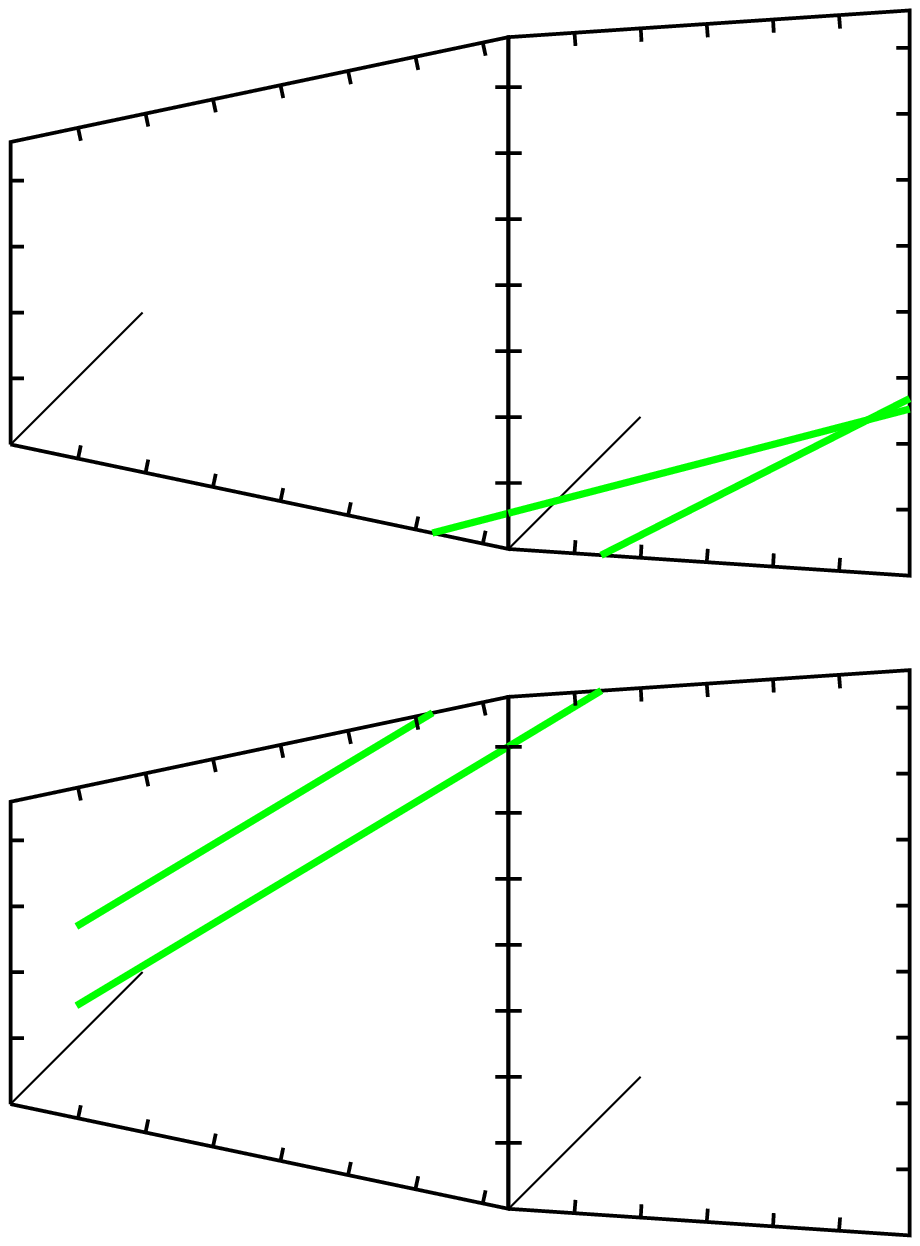}
  }
\caption{\label{fig.vertexgezeiten}
Initially parallel geodesics
that pass a vertex on opposite sides
subsequently are
no longer parallel.
In (a) the sectors are joined along the geodesic on the left
and in (b) along the geodesic on the right;
in (c) they are arranged symmetrically.
(d) Spacelike geodesics show the opposite behaviour.
(Sector model as in
figure~\protect\ref{fig.rtmodell2},
but
with double coordinate length
in time, $c\Delta t = 2.5\,\rs$.)
}
\end{figure}

\subsection{The construction of geodesics using transfer double sectors}
\label{sec.trgeodzeichnen}

\begin{figure}
  \centering
  \subfigure[]{%
    \includegraphics[scale=0.5]{\bilder/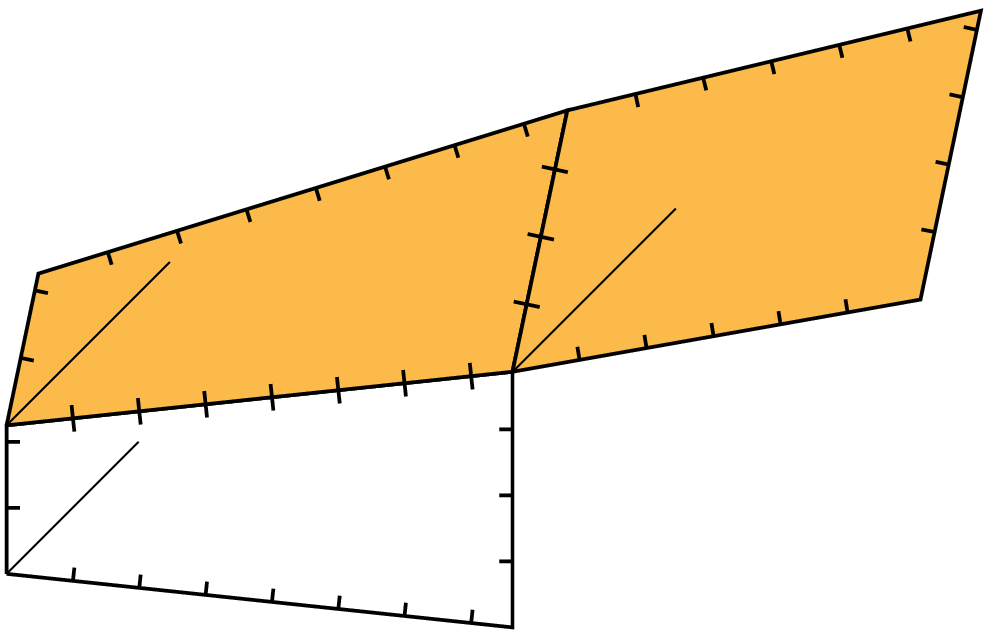}
  }\hspace*{2cm}
  \subfigure[]{%
    \includegraphics[scale=0.5]{\bilder/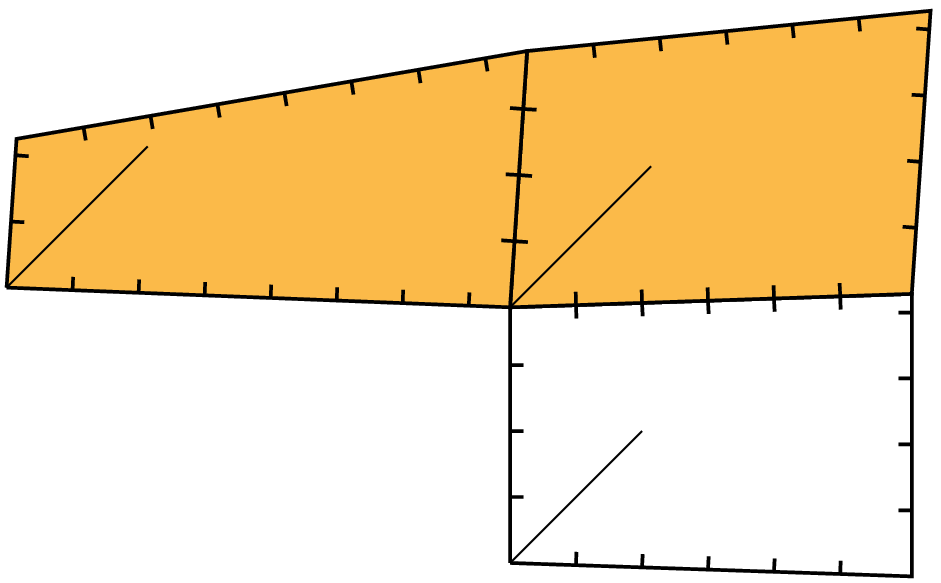}
  }
  \caption{\label{fig.trtransfer}
Transfer double sectors.
(a) For the left column of the sector model
of figure~\protect\ref{fig.rtmodell2},
(b) for the right column.
}
\end{figure}

When geodesics are constructed as described above,
in some cases
the line segment within a 
sector is very short
because it passes close to a vertex
(e.g., in figure~\ref{fig.gezeiten},
4th row from the bottom, left line).
In this case
the further construction is quite imprecise
because the subsequent direction
is determined from this short segment.
The problem can be solved
by using not a single transfer sector but a double one
(figure~\ref{fig.trtransfer}).
This is built by joining a sector of the neighbouring
column, after appropriate Lorentz transformation,
to a transfer sector.
The line on the double sector is then longer
and the construction is more precise.
In the workshops we first introduce the single transfer sectors
of figure~\ref{fig.rtmodell2}.
When the procedure is clear,
we switch to the double sectors of figure~\ref{fig.trtransfer}.

\section{Conclusions and outlook}

\label{sec.fazit3}

\subsection{Summary and pedagogical comments}

In this contribution
we have shown how paths of light and free particles
can be constructed on spacetime sector models.
The construction of null geodesics directly leads
to the phenomenon of gravitational redshift
(section~\ref{sec.rot}).
The construction of timelike geodesics
shows that describing the motion of a particle in free fall
as a geodesic in spacetime, provides
the expected up-and-down motion in space
(section~\ref{sec.wurf}).
By studying timelike geodesics of neighbouring particles,
the
connection between spacetime curvature
and Newtonian tidal forces
is revealed
(section~\ref{sec.gezeiten}).

In connection with the use of spacetime sector models
one can discuss the equivalence principle
that is expressed here in a clear way.
It states that
in sufficiently small regions of a curved spacetime
Minkowski geometry applies and that
locally all physical phenomena
are described by the special theory of relativity.
In a sector model each sector constitutes
such a small region.
The curved spacetime is explicitly made up of local regions
with Minkowski geometry.
In the sector model one can advance through curved spacetime
by passing from one Minkowski sector to the next.
The local validity of special relativity
is directly implemented on sector models,
when the world lines of light and free particles
are drawn as straight line segments within a sector.

The sector model used here
represents a 1+1-dimensional subspace
of the Schwarzschild spacetime.
It allows the construction of radial world lines.
Non-radial world lines can in principle be determined
in a 2+1-dimensional sector model,
but an implementation with models made from paper or cardboard
does not appear practicable.
An implementation using
three-dimensional interactive computer visualization
is being studied.

The workshops on redshift and particle paths presented here
were developed and tested
at Hildesheim University
in the context of an introduction to general relativity
for pre-service physics teachers
\citep{zah2013, kra2018}.
This introductory course
uses the model-based approach
described here
including the calculation of the
relevant
sector models
from their
metrics.
The calculation of sector models is introduced step by step
starting with
the sphere via the equatorial plane of a black hole
(paper~II, section 2.4)
to the spacetime of a radial ray
(section~\ref{sec.trberechnung}).
The course uses the material described in papers~I to~III
plus material from part four currently in preparation.
In the homework problems and the tutorials,
the students calculate sector models for other metrics
on their own
and use them to study curvature and geodesics.
Thus, in the model-based course
students are taught the necessary skills
for studying (to a certain extent)
the physical phenomena associated with a given metric.
Answers are here obtained graphically
that in a standard university course would be found by calculations.
An example of a problem
that can be solved with the methods of the model-based course
is the following:
`The metric of a radial ray in an expanding spacetime
is given as
$\ud s^2 = -c^2 \ud t^2 + (t/T_0)^2\, \ud x^2$,
where $T_0$ is a constant.
Two observers, each at a constant coordinate $x$, exchange light signals.
Will they observe a redshift?'
Details on the pre-service teacher course
and its evaluation are presented
by \cite{kra2018}.

Other possible uses,
e.g. in an astronomy club at school,
exist, in particular, for the workshop on redshift
because it does not require the participants to have
knowledge of special relativity.
Also, all of the material can be used
as a supplement to a mathematically oriented university course
and help to strengthen geometric insight.

\subsection{Comparison with other graphic approaches}

Sector models provide a graphic representation
of spacetime geodesics.
Other graphic representations of geodesics in spacetime
have been described using embedding surfaces
\citep{mar1999, jon2001, jon2005}.
Just as the sector models presented here,
these representations are limited to 1+1-dimensional spacetimes.
A related representation of geodesics
is the construction on so-called wedge maps
developed by \citet{dis1981}.
This construction is derived from the Regge calculus
and is carried out numerically.
The calculation is also described for 2+1-dimensional
spacetimes; light deflection and redshift are discussed.

In comparison to embedding surfaces and also to
wedge maps,
the calculation and use of sector models is
more elementary.
For a spacetime model, only a basic knowledge of special relativity
is necessary;
the determination of geodesics is carried out graphically
and the only mathematical concept that goes beyond
elementary mathematics as taught at school
is the concept of the metric.
Sector models can easily be constructed
and since they are readily duplicated,
all participants of a course can
carry out the construction of geodesics
themselves on their own models.

\subsection{Outlook}

In the model-based approach described here,
sector models are the basis for answers to
the three fundamental questions raised in paper~I
concerning
the nature of a curved spacetime,
the laws of motion,
and the relation between the distribution of matter
and the curvature of spacetime.
In paper~I curved spaces and spacetimes are represented
as sector models. In paper~II and the present contribution
geodesics are studied as paths of light and free particles.
Part four of this series will be on
the relation
between curvature and the distribution of matter.

\end{document}


\title[Sektormodelle: III. Geodäten in der Raumzeit]{%
Sektormodelle -- Ein Werkzeugkasten zur Vermittlung der
               Allgemeinen Relativitätstheorie:\\
III. Geodäten in der Raumzeit
}

\author{U Kraus und C Zahn}

\bigskip
\address{Institut f\"ur Physik, Universität Hildesheim,
Universitätsplatz 1, 31141 Hildesheim}

\eads{
\mailto{ute.kraus@uni-hildesheim.de},
\mailto{corvin.zahn@uni-hildesheim.de}
}

\begin{abstract}

{\parindent=0pt

Sektormodelle ermöglichen einen modellbasierten Zugang zur Allgemeinen
Relativitätstheorie, der auf das Verständnis der geometrischen Konzepte
abzielt und in seinen mathematischen Anforderungen
nicht über Schulmathematik hinausgeht.
Dieser Beitrag zeigt, wie die Bahnen von Licht und freien Teilchen
auf einem raumzeitlichen Sektormodell konstruiert werden können.
Als Beispiel dienen radiale Bahnen in der Nähe eines Schwarzen Lochs.
Wir beschreiben zwei Workshops zu den Themen
gravitative Rotverschiebung sowie radialer freier Fall,
die wir in dieser Form mit Bachelorstudierenden durchführen.
Der Workshop zur Rotverschiebung setzt keine Kenntnisse
der Speziellen Relativitätstheorie voraus;
in dem Workshop zum freien Fall wird die Lorentztransformation
als bekannt vorausgesetzt.
Der Beitrag beschreibt auch eine vereinfachte Berechnung des
verwendeten raumzeitlichen Sektormodells, die von Teilnehmer/innen
selbstständig durchgeführt werden kann, falls sie mit
der Minkowskimetrik vertraut sind.
Die vorgestellten Materialien stehen online unter\\
{\tt www.tempolimit-lichtgeschwindigkeit.de}
für den Unterricht zur Verfügung.
}

\end{abstract}

\maketitle

\section{Einleitung}

\label{sec.einleitung3}

\begin{figure}
  \subfigure[]{%
    \includegraphics[width=\db]{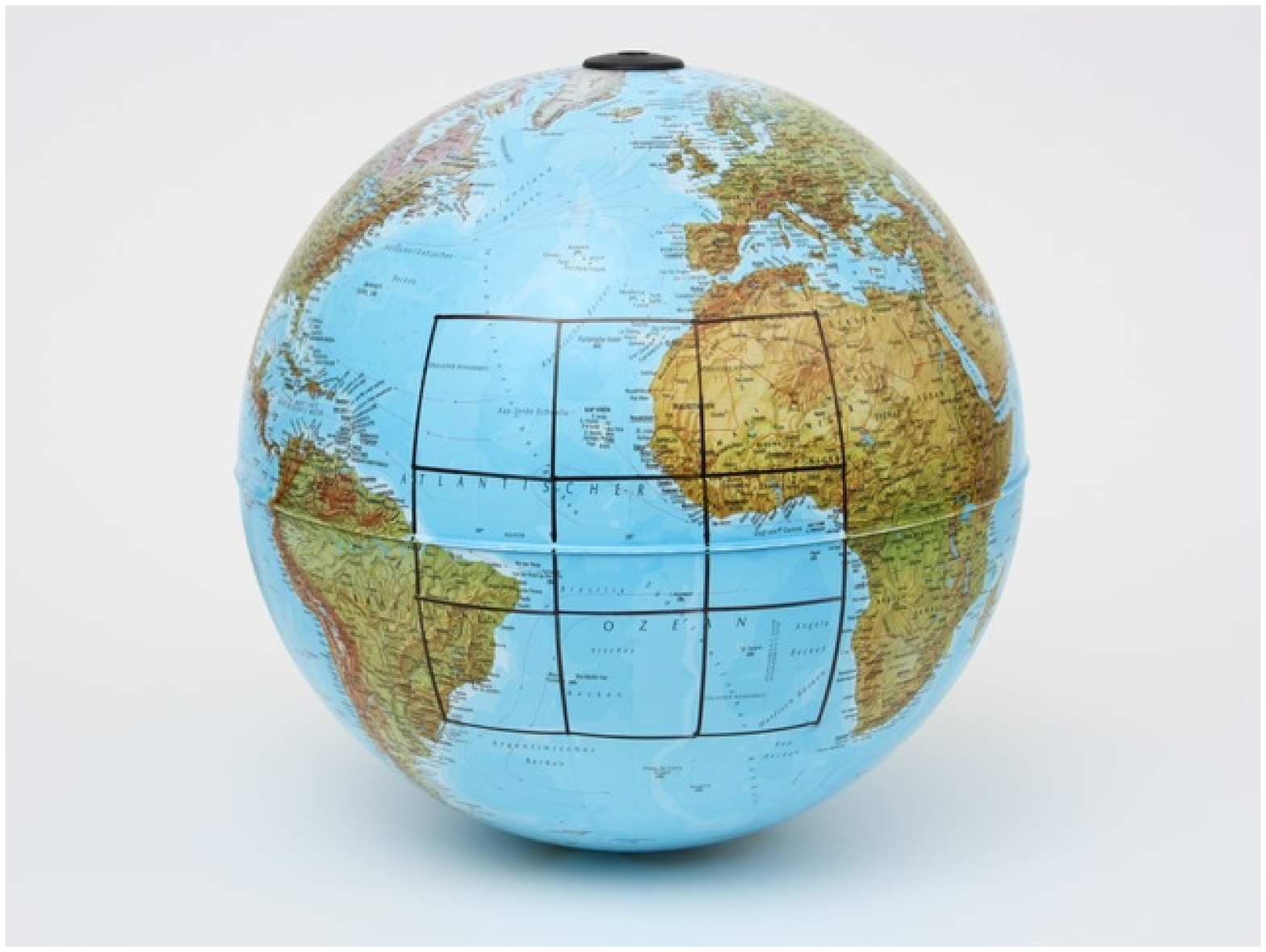}
  }\hfill%
  \subfigure[]{%
    \includegraphics[width=\db]{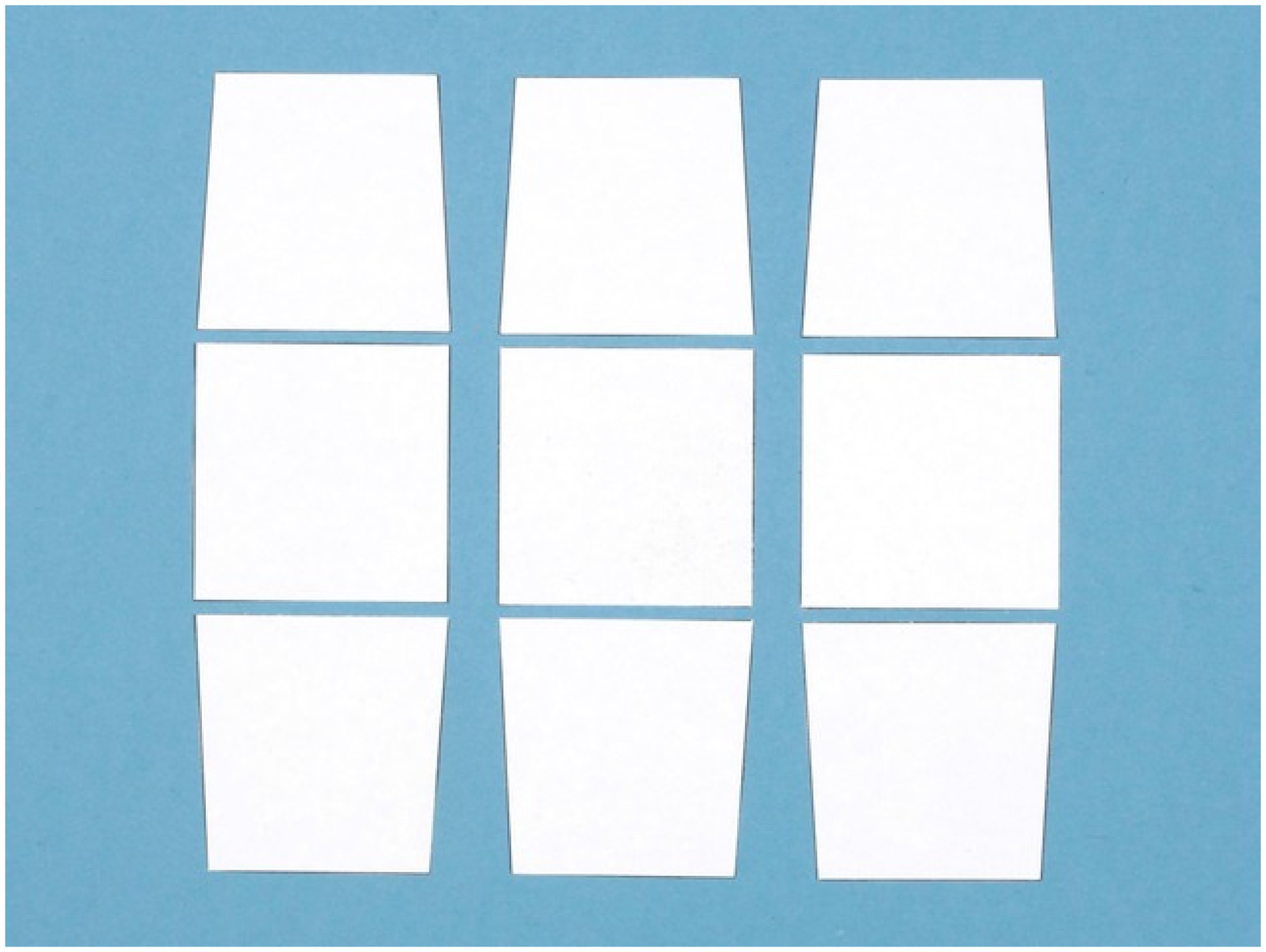}
  }\hfill%
  \subfigure[]{%
    \includegraphics[width=\db]{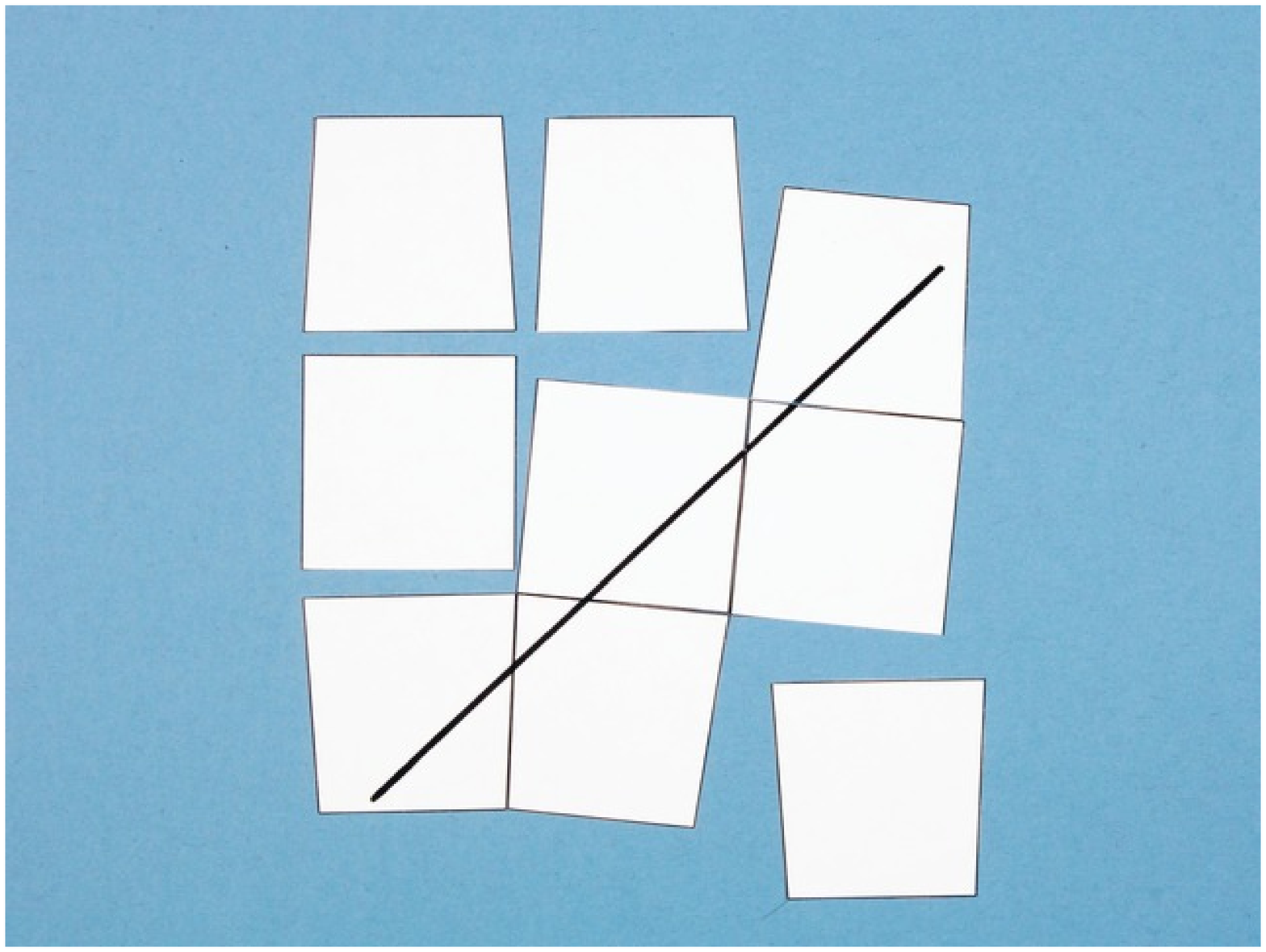}
  }
  \caption{\protect\raggedright
Sektormodell und Konstruktion von Geodäten am Beispiel der Kugeloberfläche.
Die gekrümmte Fläche wird in kleine Bereiche zerlegt (a), deren
Kantenlängen bestimmt werden. Die Sektoren werden als ebene Flächenstücke
mit denselben Kantenlängen konstruiert (b).
Eine Geodäte wird als lokal gerade Linie mit dem Lineal gezeichnet (c).
  \label{fig.zusammen3}
  }
\end{figure}

Im Hinblick auf das Ziel,
die Grundlagen der Allgemeinen Relativitätstheorie
zu vermitteln und dabei in den mathematischen Anforderungen
nicht über Schulmathematik hinauszugehen,
entwickeln wir einen Zugang,
der auf speziellen Anschauungsmodellen,
sogenannten Sektormodellen, basiert.
Dahinter steht der Grundgedanke, dass die Allgemeine
Relativitätstheorie eine geometrische Theorie ist, die deshalb
der geometrischen Anschauung zugänglich ist.
Im ersten Teil dieser Folge von Beiträgen haben wir Sektormodelle
als Anschauungsmodelle von gekrümmten Räumen und Raumzeiten
entwickelt \citep[im Folgenden als Teil~I bezeichnet]{teil1}.
Sektormodelle setzen die Darstellung gekrümmter Raumzeiten
im Regge-Kalkül
\citep{reg1961}
in Form von gegenständlichen Modellen um.
Sektormodelle können zweidimensional sein (z.\,B. eine Symmetrieebene
eines kugelsymmetrischen Sterns), dreidimensional (z.\,B. der gekrümmte
dreidimensionale Raum um ein Schwarzes Loch) oder 1+1-dimensional
(d.\,h. eine Raumzeit, in der zwei räumliche Dimensionen unterdrückt
werden, ähnlich den Minkowskidiagrammen der Speziellen Relativitätstheorie).
Das Prinzip zeigt Abb.~\ref{fig.zusammen3}
am Beispiel der Kugeloberfläche:
Die gekrümmte
Fläche wird in kleine Bereiche zerlegt,
in diesem Beispiel in Vierecke (Abb.~\ref{fig.zusammen3}(a)).
Für alle Vierecke werden die Kantenlängen bestimmt.
In der Ebene werden Vierecke mit denselben Kantenlängen konstruiert
(Abb.~\ref{fig.zusammen3}(b)).
Dies sind die Sektoren, die das Sektormodell bilden.
Das Sektormodell stellt die
gekrümmte Fläche näherungsweise dar. Die Güte der Annäherung ist durch
die Feinheit der Unterteilung bestimmt. Für didaktische Zwecke
ist eine relativ grobe Unterteilung sinnvoll.
An Sektormodellen kann man die Geometrie
der dargestellten Räume und Raumzeiten mit grafischen Methoden untersuchen.
Dazu gehört die Konstruktion von Geodäten,
die im zweiten Teil dieser Folge
\citep[im Folgenden als Teil~II bezeichnet]{teil2}
beschrieben wird.
Die Konstruktion ist eine zeichnerische Umsetzung der Bestimmung
von Geodäten im Regge-Kalkül
\citep{wil1981}.
Das prinzipielle Vorgehen zeigt Abb.~\ref{fig.zusammen3}(c).
Ausgehend von der Definition einer
Geodäte als einer lokal geraden Linie wird die Geodäte mit dem Lineal
konstruiert:
Innerhalb eines Sektors, der ja ein ebenes
Flächenstück ist, ist eine Geodäte eine gerade Linie.
Wenn die Geodäte den Rand des Sektors erreicht, wird der Nachbarsektor
angelegt und
die Linie wird über die Kante hinweg geradlinig fortgesetzt.
Sektormodelle werden maßstabsgetreu berechnet, so dass die geometrischen
Eigenschaften, die man an ihnen abliest, im Rahmen des Diskretisierungsfehlers
auch quantitativ korrekt sind.
Für Geodäten
erzielbare Genauigkeiten
werden in Teil~II untersucht.

Die Bahnen von Licht und freien Teilchen
werden in der Allgemeinen Relativitätstheorie
als Geodäten in der Raumzeit beschrieben.
Dieser Beitrag zeigt,
wie man mithilfe von Sektormodellen Geodäten in der Raumzeit
zeichnerisch bestimmen kann. Als Beispiel dienen radiale
Geodäten in der Nähe eines Schwarzen Lochs.
Wir stellen zwei Workshops vor, die wir in dieser Form mit
Bachelorstudierenden durchführen. Im Workshop zur gravitativen
Rotverschiebung
(Abschnitt~\ref{sec.workshoprot})
werden Nullgeodäten konstruiert, an denen die Rotverschiebung
verdeutlicht wird.
Der Workshop zum radialen freien Fall
(Abschnitt~\ref{sec.workshopball})
beinhaltet die Konstruktion von radialen zeitartigen Geodäten
und den Vergleich mit der Newtonschen Beschreibung des freien Falls
sowie der Gezeitenkräfte.
Fazit und Ausblick folgen in Abschnitt~\ref{sec.fazit3}.

\section{Workshop Rotverschiebung}
\label{sec.workshoprot}

In diesem Workshop werden Weltlinien von Licht
als Geodäten in der Raumzeit konstruiert.
Daran wird das Zustandekommen der
gravitativen Rotverschiebung verdeutlicht.
Als Beispiel dient die
Raumzeit eines Schwarzen Lochs,
weil die Effekte dort groß und grafisch gut darstellbar sind.
Es werden nur radiale Bahnen betrachtet;
bei der Darstellung der Raumzeit werden die anderen
Raumrichtungen unterdrückt, so dass das raumzeitliche Sektormodell
1+1-dimensional ist.

Der Workshop setzt voraus,
dass den Teilnehmer/innen
der Begriff der Geodäte
als einer lokal geraden Linie
sowie Sektormodelle zur Darstellung von Flächen mit Krümmung
bekannt sind.
Vorkenntnisse in Spezieller Relativitätstheorie
sind für diesen Workshop nicht erforderlich.
Minkowskidiagramme spielen eine Rolle und bei Bedarf
werden sie zu Beginn des Workshops im benötigten
Umfang erläutert: Es wird erstens erklärt,
dass es sich um Raum-Zeit-Diagramme handelt,
die sich von den aus der Mechanik bekannten Diagrammen
dadurch unterscheiden, dass die senkrechte Achse die
Zeitachse ist. Um die Teilnehmer/innen
mit dieser Darstellung vertraut
zu machen, zeigen wir ein Diagramm mit Weltlinien
und lassen die darin dargestellte Geschichte erzählen
(ein Beispiel steht online zur Verfügung,~\citealp{teil3}).
Zweitens wird auf die Skalierung der Achsen eingegangen,
die so gewählt wird, dass Bewegungen mit Lichtgeschwindigkeit
im Diagramm unter $45^{\circ}$ verlaufen.
Schließlich werden die Begriffe Ereignis, Weltlinie und Lichtkegel
eingeführt.

\subsection{Rotverschiebung in der Nähe eines Schwarzen Lochs}
\label{sec.rot}

\begin{figure}
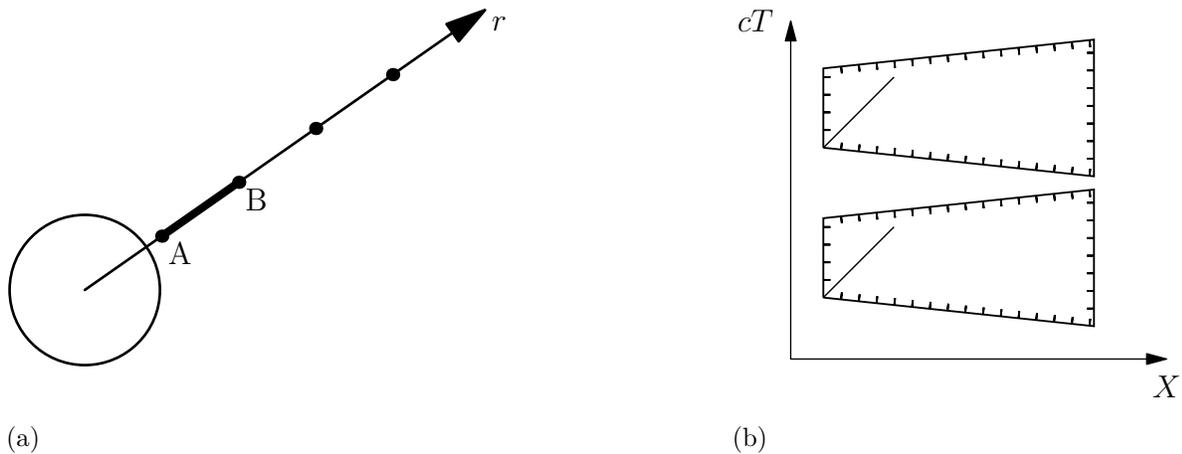

  \subfigure[]{%
    \raisebox{0.5cm}{\includegraphics{\bilder/figure2a.eps}}
  }\hfill%
  \subfigure[]{%
    \includegraphics{\bilder/figure2b.eps}
  }
\caption{\label{fig.rtmodell}
Sektormodell für die Raumzeit eines radialen Strahls
im Au\3enraum eines Schwarzen Lochs.
(a) Radialer Strahl.
(b) Raumzeitliches Sektormodell für den in (a)
kenntlich gemachten Abschnitt.
Die linke Kante stellt Ereignisse an Punkt~A dar,
die rechte Kante an Punkt~B.
Die diagonalen Linien markieren die Lichtkegel.
Das Modell kann in Zeitrichtung durch identische Sektoren
beliebig erweitert werden.
}
\end{figure}
Der Workshop beginnt mit der Erläuterung,
dass die Allgemeine Relativitätstheorie
die Bahnen von Licht und freien Teilchen
als Geodäten in der Raumzeit beschreibt.
Es wird dann ein Sektormodell vorgestellt,
das die Raumzeit eines radialen Strahls
im Außenraum eines Schwarzen Lochs
darstellt.
Die Teilnehmer/innen können das Sektormodell selbst berechnen
(Abschnitt~\ref{sec.trberechnung})
oder eine Vorlage verwenden
(online verfügbar, \citealp{teil3}).
Im Gedankenexperiment entsteht das Sektormodell aus Messungen
in der Nähe eines Schwarzen Lochs:
Astronauten reisen zu dem Schwarzen Loch und positionieren
sich dort längs eines radialen Strahls.
Sie wählen eine Reihe von Ereignissen aus und
nutzen sie als Eckpunkte um die Raumzeit des radialen Strahls
in Vierecke einzuteilen.
Zur Definition eines einzelnen Vierecks
werden zwei Positionen auf dem radialen Strahl ausgewählt
(Abb.~\ref{fig.rtmodell}(a)).
Zwei Ereignisse an der inneren
sowie zwei an der äußeren Position
bilden die vier Eckpunkte.
Der viereckige Ausschnitt der gekrümmten Raumzeit
wird
durch einen Sektor eines Minkowskiraums
dargestellt
(Abb.~\ref{fig.rtmodell}(b));
die Gesamtheit der Sektoren bildet das Sektormodell.%
\footnote{Das Sektormodell überdeckt den Bereich von $1,25$ bis
$2,5$ Schwarzschildradien
in der Schwarzschildschen Radialkoordinate,
s. Abschnitt~\ref{sec.trberechnung}.}
Da die Raumzeit des Schwarzen Lochs statisch ist
(wir betrachten ein nichtrotierendes Schwarzes Loch),
kann man eine Darstellung wählen, in der die Form der Sektoren
von der Zeit unabhängig ist (s.~Abschnitt~\ref{sec.trberechnung}).
Dies wurde hier umgesetzt,
so dass das Sektormodell in Zeitrichtung
durch weitere identische Sektoren beliebig erweitert werden kann.

\begin{figure}
\begin{center}
\includegraphics{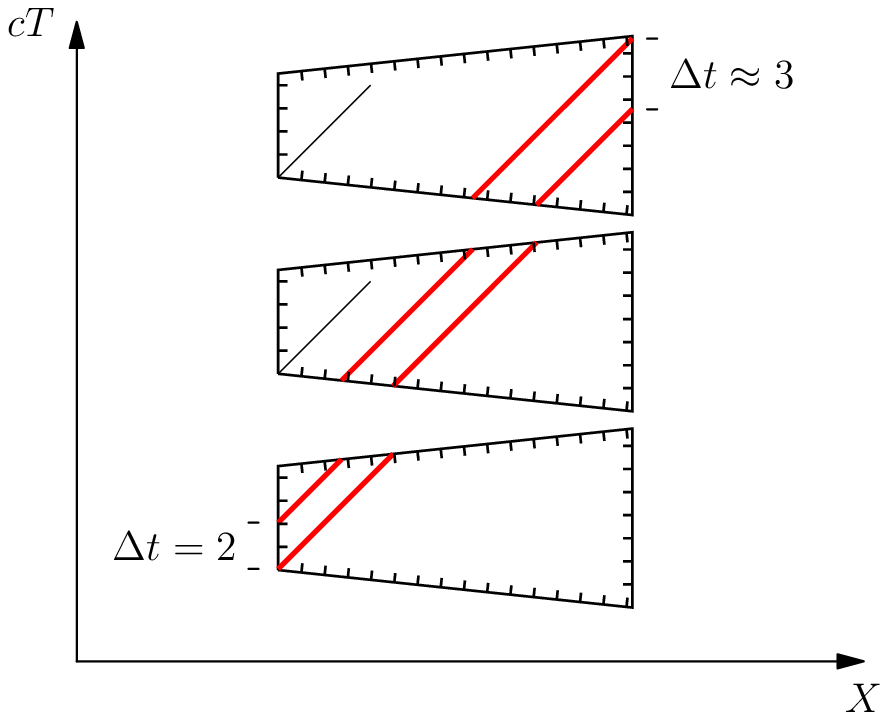}
\end{center}
\caption{\label{fig.rot}
Weltlinien zweier Lichtsignale,
die radial nach außen laufen.
Ein Beobachter am Innenrand der Sektorspalte (links)
sendet die Signale im Abstand von zwei Zeiteinheiten aus.
Ein Beobachter am Außenrand (rechts) empfängt sie im Abstand von
ca.~drei Zeiteinheiten.
}
\end{figure}

Auf dem Sektormodell wird als erstes
die Weltlinie eines Lichtsignals konstruiert,
das radial nach außen läuft.
Ausgehend von dem Startpunkt in der linken unteren Ecke des Modells
wird eine lokal gerade Linie gezeichnet,
die in Richtung des Lichtkegels verläuft
(Abb.~\ref{fig.rot}, untere Linie).
Innerhalb eines Sektors ist die Linie gerade.
Erreicht man beim Zeichnen den Rand,
so wird die Linie im Nachbarsektor fortgesetzt:
Die Position am Rand wird
auf den entsprechenden Rand des Nachbarsektors übertragen;
zur Unterstützung tragen die Kanten äquidistante Markierungen.
Die Richtung im Nachbarsektor ist, da es sich um eine Weltlinie
von Licht handelt, durch den Lichtkegel vorgegeben.\footnote{%
Alternativ kann man den Nachbarsektor anlegen und die Geodäte
geradlinig fortsetzen, wie in Abb.~\protect\ref{fig.zusammen3}
beschrieben. Das Vorgehen beim Anlegen von raumzeitlichen Sektoren
wird in Abschnitt~\ref{sec.zeitartig} erläutert.}

Im zweiten Schritt wird die Übermittlung
von zwei aufeinanderfolgenden Lichtsignalen untersucht.
Ein Beobachter, der sich nahe am Schwarzen Loch
in einer festen Entfernung
(Punkt~A am Innenrand der Sektorspalte) befindet,
sendet zwei
Lichtsignale in kurzem zeitlichem Abstand
nach außen.
In Abb.~\ref{fig.rot} beträgt dieser Abstand
zwei Zeiteinheiten.
Ein zweiter Beobachter, der sich weiter außen
ebenfalls in fester Entfernung zum Schwarzen Loch befindet
(Punkt~B am Außenrand der Sektorspalte),
empfängt die beiden Signale.
Um festzustellen,
in welchem zeitlichen Abstand
der äußere Beobachter die Signale erhält,
wird
die Weltlinie des zweiten Signals
hinzugefügt (Abb.~\ref{fig.rot}, obere Linie).
Man erkennt,
dass der zeitliche Abstand am Außenrand der Spalte
rund drei Zeiteinheiten beträgt.
Wenn man die beiden Signale
als aufeinanderfolgende Wellenberge
einer elektromagnetischen Welle deutet,
dann folgt aus der Konstruktion,
dass die Welle außen mit einer vergrößerten Periode empfangen wird.
Strahlung, die sich vom Schwarzen Loch
entfernt, wird also rotverschoben.
Das Verhältnis der Perioden
$P_{\rm außen}$ und $P_{\rm innen}$ am äußeren Punkt~B
bzw. am inneren Punkt~A
ergibt sich aus der
Konstruktion zu
$P_{\rm außen} / P_{\rm innen} \approx 1,5$.
Der berechnete exakte Wert ist
$P_{\rm außen} / P_{\rm innen}
   = \sqrt{(1-\rs/r_{\rm außen})/(1-\rs/r_{\rm innen})}
   = 1,73$
wobei
$r_{\rm innen} = 1,25\,\rs$
und
$r_{\rm außen} = 2,5\,\rs$
die Radialkoordinaten der Punkte~A und~B sind.
Der grafisch bestimmte Wert ist 13\% zu klein;
diese Abweichung ist der relativ groben
Auf"|lösung des Sektormodells geschuldet.

\subsection{Berechnung des raumzeitlichen Sektormodells}
\label{sec.trberechnung}

Eine vereinfachte Berechnung von Sektormodellen
wird in Teil~II (Abschnitt~2.4) eingeführt
und hier auf eine 1+1-dimensionale
gekrümmte Raumzeit übertragen.
Diese Berechnung setzt die Kenntnis der Minkowskimetrik
voraus.
Mit dem vereinfachten Verfahren kann
das Sektormodell
von den Teilnehmer/innen des Workshops
mit Mitteln der Schulmathematik
selbst berechnet werden.
Dies versetzt sie in die Lage,
auch andere gekrümmte Raumzeiten
ausgehend von
deren Metrik
mit Hilfe von Sektormodellen zu untersuchen.
Die mit dem vereinfachten Verfahren verbundenen
Näherungen werden in Teil~II diskutiert.

Die Konstruktion geht von der Metrik
\begin{equation}
\label{eq.ssm-rt}
\ud s^2 = - \left (1 - \frac{\rs}{r} \right) c^2 \ud t^2 +
            \frac{1}{1 - \rs/r } \ud r^2
\end{equation}
aus,
wobei $t$ und $r$ die üblichen Schwarzschildkoordinaten
sind und $\rs = 2 G M / c^2$
der Schwarzschildradius der Zentralmasse $M$ mit
der Newtonschen Gravitationskonstanten $G$
und der Lichtgeschwindigkeit $c$.

\begin{figure}
  \centering
  \subfigure[]{%
    \includegraphics{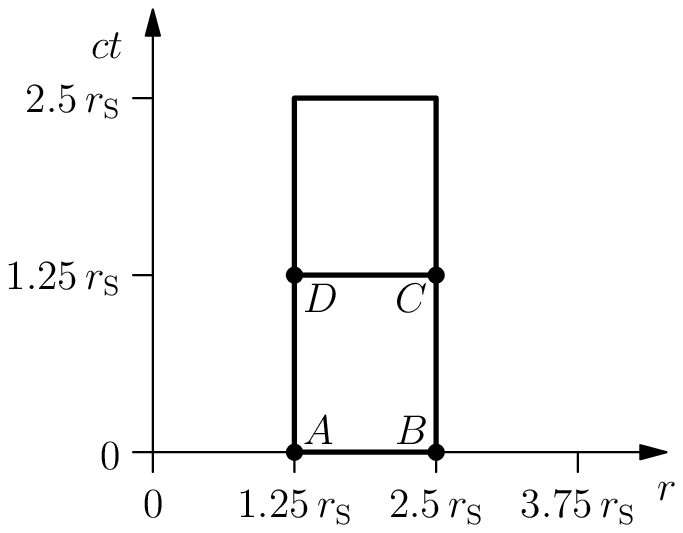}
  }\hfil%
  \subfigure[]{%
    \includegraphics{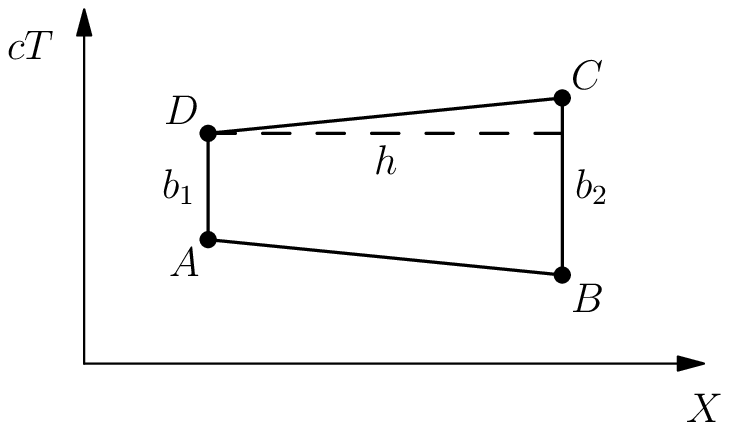}
  }
  \caption{\label{fig.rtteilung}
Zur Berechnung des raumzeitlichen Sektormodells eines
radialen Strahls.
(a) Die Aufteilung der Raumzeit im Koordinatenraum.
(b) Jeder Sektor wird als symmetrisches Trapez konstruiert.
}
\end{figure}

Das Sektormodell stellt räumlich einen Abschnitt eines radialen
Strahls zwischen $r=1,25\, \rs$ und $r=2,5\, \rs$ dar,
der die Koordinatenlänge
$\Delta r = 1,25\, \rs$ hat.
Die Zeitkoordinate $t$ wird in Abschnitte der Länge $\Delta t$
mit $c\Delta t = 1,25\, \rs$ eingeteilt
(Abb.~\ref{fig.rtteilung}(a)).
Da die Metrik von der Zeitkoordinate nicht abhängt,
braucht nur ein einziger zeitlicher Abschnitt berechnet zu werden.

Die Berechnung der Kantenintervalle ergibt
für die zeitartigen Kanten
\begin{equation}
  \Delta s_t^2 (r) = - \left (1 - \frac{\rs}{r} \right) c^2 \Delta t^2
     \hspace*{1cm} (\Delta r = 0)
\end{equation}
und
für die raumartigen Kanten
\begin{equation}
\Delta s_r^2 = \frac{1}{(1-\rs/\rrmi)}\, \Delta r^2
     \hspace*{1cm} (\Delta t = 0)
,
\end{equation}
wobei
der Metrikkoeffizient
für die mittlere Koordinate
$\rrmi=(r_1+r_2)/2$ berechnet wird,
mit den Koordinaten $r_1$ und
$r_2$ der zugehörigen Eckpunkte.
Da in Zeitrichtung identische Sektoren aneinandergereiht werden,
wird jeder Sektor zeitsymmetrisch als Trapez konstruiert:
Gemäß Abb.~\ref{fig.rtteilung}(b)
werden die Grundseiten in Zeitrichtung gelegt;
sie haben die Längen
$b_1=\sqrt{-\Delta s_t^2 (r=1,25\,\rs)}$
und $b_2=\sqrt{-\Delta s_t^2 (r=2,5\,\rs)}$.
Die Höhe $h$ des Trapezes wird so bestimmt,
dass die Schenkel das Intervall
$\Delta s_r^2$ haben (Abb.~\ref{fig.rtteilung}(b)):
\begin{equation}
\Delta s_r^2
= - \left ( \frac{b_2 - b_1}{2} \right )^2 + h^2.
\end{equation}
Das Ergebnis ist der in Abb.~\ref{fig.rtmodell} gezeigte Sektor.

\section{Workshop Teilchenbahnen}
\label{sec.workshopball}

In diesem Workshop werden Weltlinien von
frei fallenden Teilchen in der Nähe
eines Schwarzen Lochs konstruiert.
Wie im vorangegangenen Abschnitt werden nur
radiale Bahnen betrachtet, so dass ein 1+1-dimensionales
raumzeitliches Sektormodell verwendet werden kann.
Anhand der Teilchenbahnen wird der Zusammenhang
zwischen der
relativistischen und der
klassischen Beschreibung
von Bewegung im Schwerefeld verdeutlicht.
Der Workshop setzt voraus,
dass die Teilnehmer/innen mit der Lorentztransformation vertraut sind.

\begin{figure}
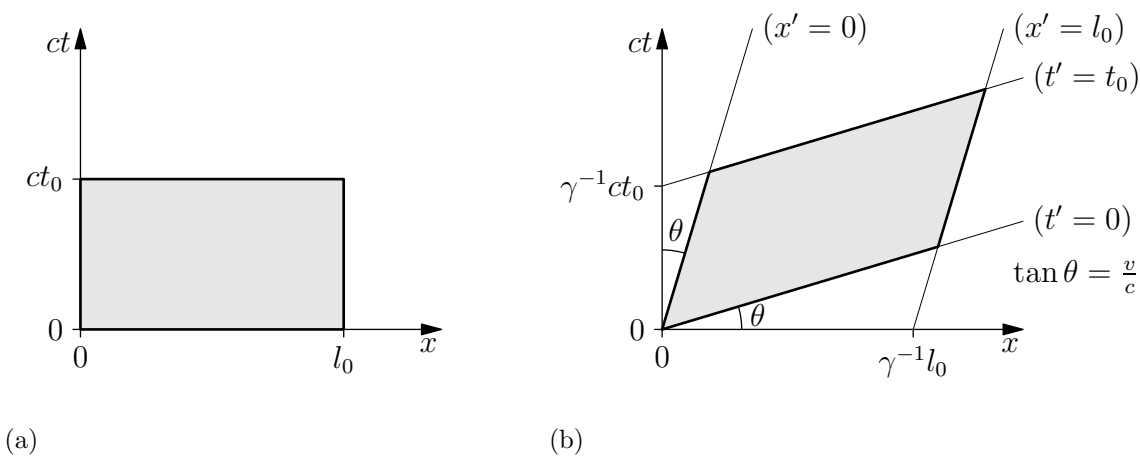

  \centering
  \subfigure[]{%
    \includegraphics{\bilder/figure5a}
  }\hfil%
  \subfigure[]{%
    \includegraphics{\bilder/figure5b}
  }
\caption{\label{fig.raumschiff}
Grafische Darstellung eines Raumzeitsektors in zwei verschiedenen
Bezugssystemen.
Die Ereignisse liegen in einem Raumschiff der
Länge $l_0$ bei Bordzeiten zwischen null und $t_0$.
(a) Darstellung im Ruhesystem des Raumschiffs.
(b)
Darstellung im
Ruhesystem einer Raumstation, an der sich das Raumschiff mit
Geschwindigkeit $v=0,3\,c$
vorbeibewegt ($\gamma=1 / \sqrt{1-v^2/c^2}$).
}
\end{figure}

\subsection{Die Konstruktion zeitartiger Geodäten}
\label{sec.zeitartig}

Um die Bahnen frei fallender Teilchen in der Nähe eines
Schwarzen Lochs zu untersuchen, werden ihre Weltlinien
als Geodäten
auf einem Sektormodell konstruiert.
Wie in den bisherigen Beispielen
werden die Geodäten innerhalb eines Sektors als gerade Linien
gezeichnet und beim Erreichen einer Kante in den Nachbarsektor
fortgesetzt.
Anders als bei den
in Abschnitt~\ref{sec.workshoprot}
betrachteten Nullgeodäten
ist die Richtung im Nachbarsektor
aber nicht durch den Lichtkegel vorgegeben.
Man muss also 
den Nachbarsektor anlegen
und die Linie über die gemeinsame Kante hinweg
geradlinig fortsetzen.
Das Anlegen ist im raumzeitlichen Fall komplizierter
als im rein räumlichen.
Man sieht leicht ein,
dass es nicht damit getan wäre,
den Nachbarsektor auszuschneiden
und in Position zu drehen:
Da die Lichtgeschwindigkeit im Nachbarsektor
denselben Wert hat wie im Startsektor,
müssen die Lichtkegel der beiden Sektoren zusammenfallen.
Dies legt die Orientierung des Nachbarsektors fest.

Im Workshop verwenden wir
ein konkretes Beispiel für einen Raumzeitsektor
um zu verdeutlichen,
auf welche Weise das Anlegen
möglich ist.
Wir betrachten
ein langes, sehr dünnes Raumschiff mit Ruhelänge $l_0$.
Der Raumzeitsektor soll aus denjenigen Ereignissen bestehen,
die innerhalb des Raumschiffs liegen und
Bordzeiten zwischen null und $t_0$ haben.
Die Teilnehmer/innen zeichnen diesen Raumzeitsektor
in ein Minkowskidiagramm ein,
zunächst im Bezugssystem des Raumschiffs (Abb.~\ref{fig.raumschiff}(a)),
anschlie\3end im Bezugssystem
einer Raumstation, an der sich das Raumschiff mit konstanter
Relativgeschwindigkeit
vorbeibewegt
(Abb.~\ref{fig.raumschiff}(b)).
Die gezeichnete Form des Sektors,
d.\,h.\ seine im euklidischen Sinne verstandene geometrische Form,
hängt offensichtlich vom Bezugssystem ab.
Abb.~\ref{fig.raumschiff}(a) und~(b) sind zwei verschiedene
grafische Darstellungen
ein und desselben Raumzeitsektors.
Sie gehen durch eine Lorentztransformation auseinander hervor.

\begin{figure}
  \centering
  \includegraphics[scale=0.5]{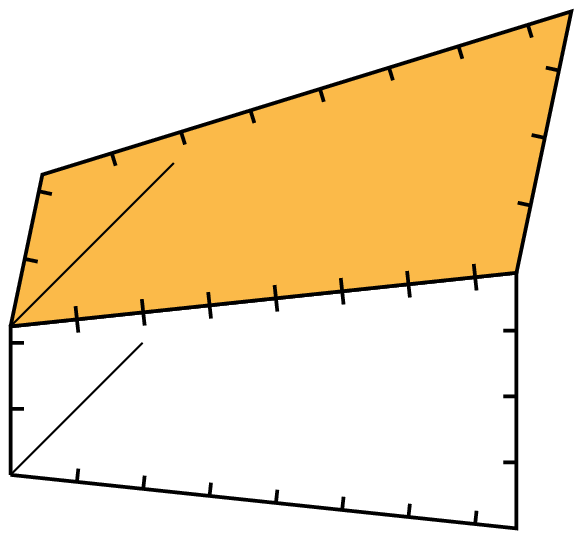}
  \caption{\label{fig.trtransfersektor}
Anlegen eines raumzeitlichen Sektors
im Sektormodell von Abb.~\protect\ref{fig.rtmodell}.
Der obere Sektor
wurde so transformiert,
dass er an den unteren Sektor angelegt werden kann
($v/c = 0,21$).
}
\end{figure}

\begin{figure}
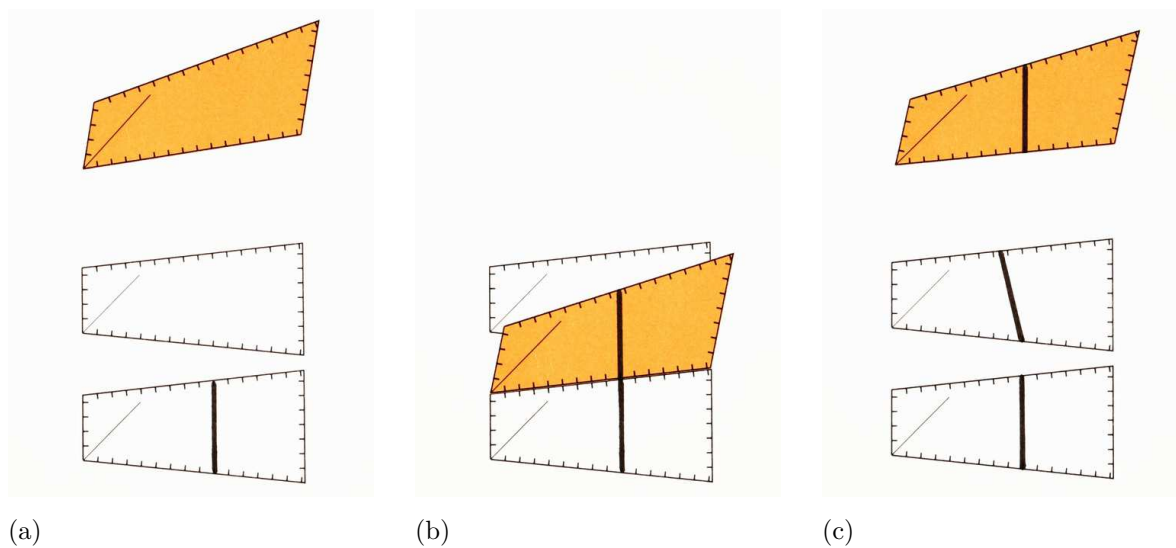

  \centering
  \subfigure[]{%
    \includegraphics[width=\db]{\bilder/figure7a.eps}
  }\hfill%
  \subfigure[]{%
    \includegraphics[width=\db]{\bilder/figure7b.eps}
  }\hfill%
  \subfigure[]{%
    \includegraphics[width=\db]{\bilder/figure7c.eps}
  }
  \caption{\label{fig.trgeodzeichnen}
Konstruktion einer Geodäte auf dem raumzeitlichen Sektormodell.
(a) Ein lorentztransformierter Sektor
dient als Transfersektor (farbig markiert).
(b) Die Geodäte wird geradlinig auf den Transfersektor fortgesetzt.
(c) Die Linie wird vom Transfersektor auf den Nachbarsektor in
der ursprünglichen, symmetrischen Form übertragen.
}
\end{figure}

Dies ermöglicht das Anlegen von Sektoren,
da man den Nachbarsektor
durch eine Lorentztransformation mit der geeigneten Geschwindigkeit
in die passende Form bringen kann
(Abb.~\ref{fig.trtransfersektor}).
Die Transformation, die das Anlegen des Nachbarsektors ermöglicht,
ist also im räumlichen Fall eine Rotation, im raumzeitlichen Fall
eine Lorentztransformation.

Beim Zeichnen von Geodäten
ist es zweckmäßig,
einen transformierten Sektor als Transfersektor
(Teil~II, Abschnitt 2.3) zu benutzen:
Eine Geodäte, die den Rand eines Sektors erreicht
(Abb.~\ref{fig.trgeodzeichnen}(a)),
wird über die Kante geradlinig auf den angelegten Transfersektor
fortgesetzt (Abb.~\ref{fig.trgeodzeichnen}(b))
und anschließend auf den Nachbarsektor in der symmetrischen Form
übertragen (Abb.~\ref{fig.trgeodzeichnen}(c)).
Dieser Übertrag macht die Lorentztransformation rückgängig.
Gerade Linien werden dabei wieder in gerade Linien transformiert.
Man überträgt anhand der Randmarkierungen die Endpunkte der Linie
und verbindet sie geradlinig (Abb.~\ref{fig.trgeodzeichnen}(c)).

\begin{figure}
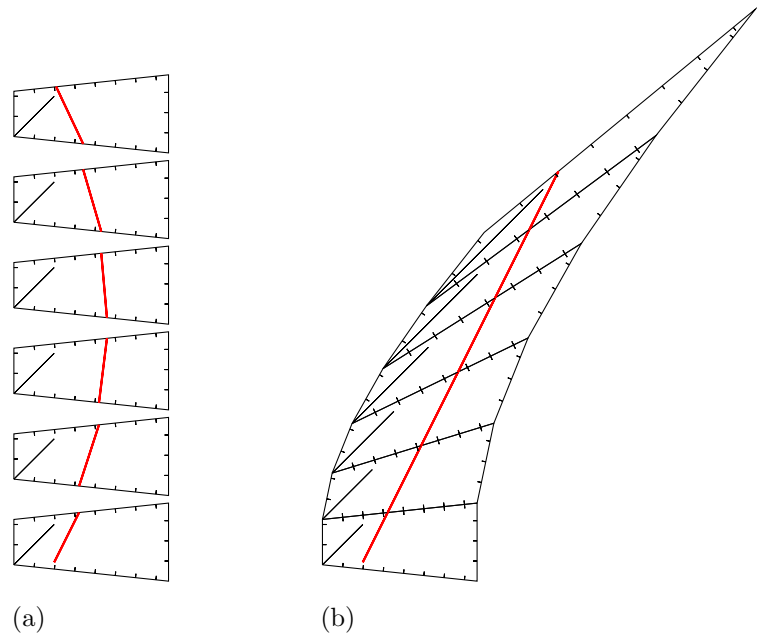

  \centering
  \subfigure[]{%
    \includegraphics[scale=0.4]{\bilder/figure8a.eps}
  }\hspace*{2cm}%
  \subfigure[]{%
    \includegraphics[scale=0.4]{\bilder/figure8b.eps}
  }
  \caption{\label{fig.wurf}
Senkrechter Wurf.
(a) Auf dem Sektormodell konstruierte Geodäte.
(b) Werden alle
   Sektoren aneinandergefügt,
   so ist der geradlinige Verlauf der Weltlinie offensichtlich.
}
\end{figure}

\subsection{Senkrechter Wurf}
\label{sec.wurf}

Ein Teilchen wird in der Nähe eines Schwarzen Lochs
senkrecht nach oben geworfen. Gesucht ist seine Bahn.
Anschaulich ist klar, dass das Teilchen
eine maximale H\"ohe erreicht und anschlie\3end
zur\"uckf\"allt
(eine Startgeschwindigkeit kleiner als
die Fluchtgeschwindigkeit vorausgesetzt).

In der relativistischen Beschreibung folgt das Teilchen,
da es sich im freien Fall befindet, einer Geod\"aten,
d.\,h. seine Weltlinie ist lokal gerade.
Wie sind diese beiden Aussagen
 -- geradlinige Weltlinie einerseits
und Auf-/Abbewegung andererseits -- miteinander vereinbar?

Zur Konstruktion der Weltlinie
wird das Sektormodell aus Abb.~\ref{fig.rtmodell}
mit sechs Zeilen verwendet
sowie ein passend transformierter
Transfersektor
(Abb.~\ref{fig.trtransfersektor}).
Nach der Vorgabe eines Startorts und einer zeitartigen,
nach außen weisenden Startrichtung
wird die Weltlinie als Geodäte auf das Sektormodell
gezeichnet (Abb.~\ref{fig.wurf}(a)):
Die als lokal gerade Linie konstruierte Geod\"ate erreicht
zun\"achst gr\"o\3ere Abstände vom Schwarzen Loch
und dann wieder kleinere.
Die raumzeitliche Geod\"ate liefert also die erwartete
Auf- und Abbewegung im Raum.
Ergänzend sind in
Abb.~\ref{fig.wurf}(b) alle
Sektoren
aneinander angefügt,
so dass der geradlinige Verlauf der Linie offensichtlich ist.
Um die Geodäte
wie in dieser Abbildung
am Stück zu zeichnen,
benötigt man etliche,
mit verschiedenen Geschwindigkeiten
lorentztransformierte Darstellungen des Sektors.
Die Konstruktion auf dem Sektormodell
zeigt sowohl die gerade Linie in der Raumzeit
als auch die Auf-/Abbewegung im Raum und
macht so den Zusammenhang zwischen beiden völlig transparent.

\begin{figure}
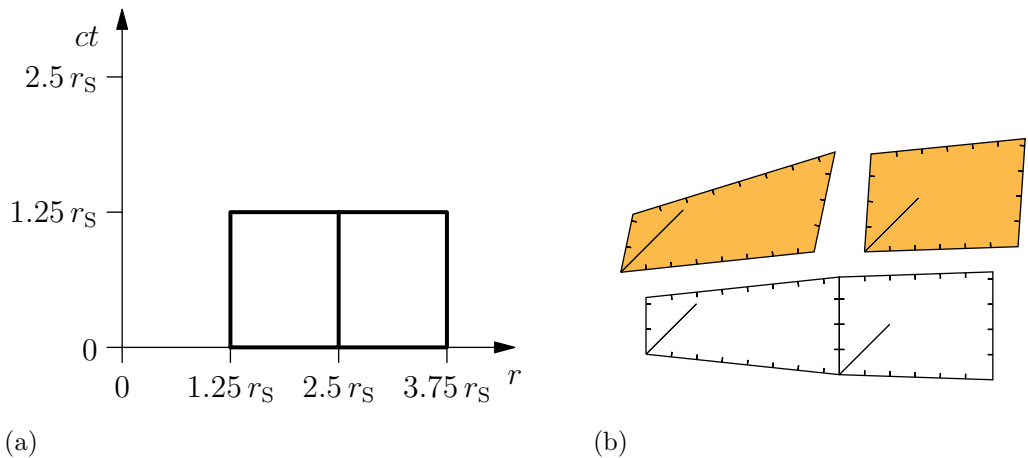

  \centering
  \subfigure[]{%
    \includegraphics{\bilder/figure9a.eps}
  }\qquad
  \subfigure[]{%
    \includegraphics[scale=0.5]{\bilder/figure9b.eps}
  }
  \caption{\label{fig.rtmodell2}
Raumzeitliches Sektormodell
eines radialen Strahls im Außenraum eines Schwarzen Lochs.
(a) Aufteilung der Raumzeit im Koordinatenraum.
(b) Sektoren in symmetrischer Form (unten) und geeignet
   lorentztransformierte Transfersektoren (oben).
Dies ist eine Erweiterung des
in Abb.~\protect\ref{fig.rtmodell} gezeigten Modells um
eine zweite Spalte mit zugehörigem Transfersektor
(rechts, $v/c = 0,067$).
Das Modell kann in Zeitrichtung durch identische Zeilen erweitert
werden.
}
\end{figure}

\subsection{Gezeitenkräfte und die Krümmung der Raumzeit}
\label{sec.gezeiten}

Wenn man zusätzlich zu diesem Workshop
auch den in Teil~I beschriebenen Workshop über Krümmung
durchführt, kann man
anhand von Geodäten
die physikalische Bedeutung der raumzeitlichen Krümmung
verdeutlichen.
Hierfür wird das Sektormodell
aus Abb.~\ref{fig.rtmodell}
um eine zweite Spalte erweitert
(Abb.~\ref{fig.rtmodell2}),
so dass es den radialen Strahl im Bereich $1,25\, \rs$ bis
$3,75\, \rs$ in zwei Spalten überdeckt.
Es werden acht Zeilen des Modells
und ein Transfersektor je Spalte
benutzt.

In einem lokalen Inertialsystem,
das sich relativ zum Schwarzen Loch momentan in Ruhe befindet,
werden zwei in radialer Richtung leicht
versetzte
Teilchen
betrachtet.
Sie werden gleichzeitig und aus der Ruhe losgelassen,
so dass sie hintereinander radial in das Schwarze Loch fallen.
In der klassischen Beschreibung nimmt die Schwerkraft nach außen ab,
so dass das
äußere Teilchen zu jedem Zeitpunkt eine geringere Beschleunigung
erfährt als das
innere.
Die beiden frei fallenden Teilchen sind deshalb relativ zueinander
beschleunigt:
Gezeitenkräfte bewirken, dass die Relativgeschwindigkeit
der Teilchen anwächst.

\begin{figure}
\centering
  \includegraphics[scale=0.35]{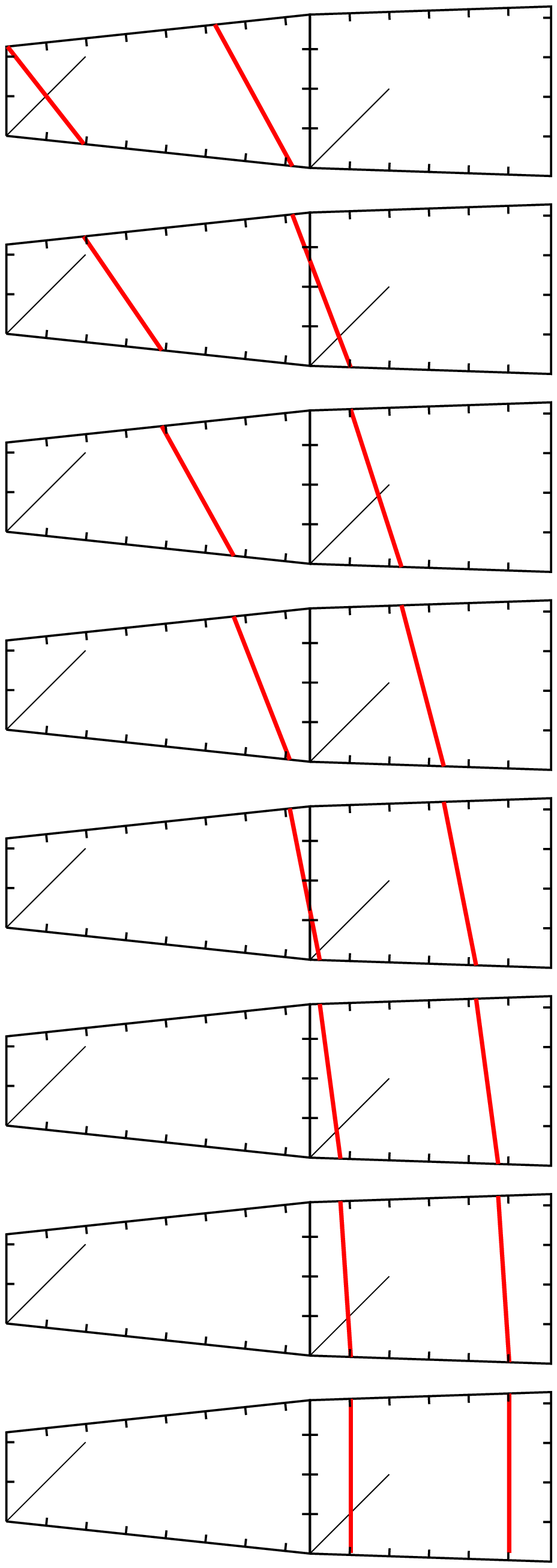}
  \caption{\label{fig.gezeiten}
 Die Weltlinien zweier Teilchen, die gleichzeitig aus der Ruhe losgelassen
 werden und
 hintereinander
 auf ein Schwarzes Loch zu fallen.
}
\end{figure}

In der Beschreibung der Allgemeinen Relativitätstheorie
sind die Weltlinien der beiden Teilchen Geodäten, die anfangs
parallel verlaufen.
Diese Geodäten
werden
auf dem Sektormodell konstruiert
(Abb.~\ref{fig.gezeiten}).
Beide Weltlinien starten
in Richtung der lokalen Zeitachse
(Abb.~\ref{fig.gezeiten}, unterste Zeile).
Die anfänglich parallelen Weltlinien laufen zunehmend auseinander,
d.\,h.
es tritt eine Relativgeschwindigkeit auf, die anwächst.
Die Konstruktion verdeutlicht,
wie die Relativbeschleunigung
zustandekommt:
Die Weltlinien sind solange parallel, bis sie das erste Mal
an einem Vertex auf verschiedenen Seiten vorbeilaufen
(Abb.~\ref{fig.gezeiten}, 4./5. Zeile von unten).
An jedem weiteren Vertex, der zwischen den Weltlinien liegt,
vergrößert sich der Unterschied in den Richtungen
und damit die Relativgeschwindigkeit.

Abb.~\ref{fig.vertexgezeiten} zeigt den Verlauf
um einen einzelnen Vertex genauer.
Hier werden der Deutlichkeit halber Sektoren verwendet,
die in Zeitrichtung die doppelte Koordinatenlänge
haben ($c \Delta t = 2,5\,\rs $).
Die Sektoren sind in
den Abbildungen~\ref{fig.vertexgezeiten}(a) und~(b) längs
der linken bzw. längs der rechten Geodäte aneinandergelegt
(wobei die obere Zeile als Ganzes jeweils passend lorentztransformiert ist);
in
Abb.~\ref{fig.vertexgezeiten}(c)
sind die Sektoren symmetrisch angeordnet.

Wie in Teil~I beschrieben, geben in einem Sektormodell
die sogenannten Defizitwinkel der Vertizes die Krümmung an.
Der Defizitwinkel des hier betrachteten Vertex
erscheint in den Abbildungen~\ref{fig.vertexgezeiten}(a) und~(b),
in denen die vier Sektoren
um den gemeinsamen Vertex
bestmöglich
zusammengelegt sind,
als verbleibende Lücke.
Dieser Defizitwinkel
liegt in raumartiger Richtung und ist positiv\footnote{%
Der Defizitwinkel ist positiv, wenn
nach Zusammenschieben aller an einen Eckpunkt angrenzenden
Sektoren eine keilförmige Lücke verbleibt.
Er ist negativ, wenn
nach Zusammenschieben aller an einen Eckpunkt angrenzenden
Sektoren bis auf einen die verbleibende Lücke
zu klein ist für den letzten Sektor.};
mit der hier verwendeten Signatur ist nach Konvention
die raumzeitliche Krümmung positiv.
Nach Konstruktion hängt der Winkel,
den die Richtungen der anfangs parallelen Geodäten
nach dem Vertex einschließen,
mit dem Defizitwinkel zusammen.
Die Abbildungen~\ref{fig.gezeiten} und~\ref{fig.vertexgezeiten} zeigen also,
dass eine positive raumzeitliche Krümmung
mit dem Auseinanderlaufen
von anfangs parallelen benachbarten Weltlinien
verbunden ist;
das Umgekehrte gilt für den Fall negativer Krümmung.
Die raumzeitliche Krümmung ist der Anschauung weniger gut
zugänglich als die räumliche.
Der Verlauf benachbarter Weltlinien liefert
aber ein anschauliches Kriterium.
Die Konstruktion zeigt damit auf,
wie die Relativbeschleunigung der beiden Teilchen
in der relativistischen Beschreibung zustandekommt.
Sie lässt sich auf die Defizitwinkel an den Vertizes
und damit auf die Krümmung zurückführen.
Hieran wird die physikalische Bedeutung der raumzeitlichen Krümmung
deutlich: Sie entspricht der Newtonschen Gezeitenkraft.

Ergänzend zeigt Abb.~\ref{fig.vertexgezeiten}(d)
das Verhalten anfänglich paralleler raumartiger Geodäten
an demselben
Vertex: Sie laufen nach dem Vertex aufeinander zu.
Das gegenläufige Verhalten von zeitartigen und raumartigen Geodäten
spiegelt das entsprechende Verhalten der Defizitwinkel
in zeitartiger bzw. in raumartiger Richtung wider
(s.~Teil~I, Abschnitt~4).
\newcommand{\figelfscale}{0.30}

\begin{figure}
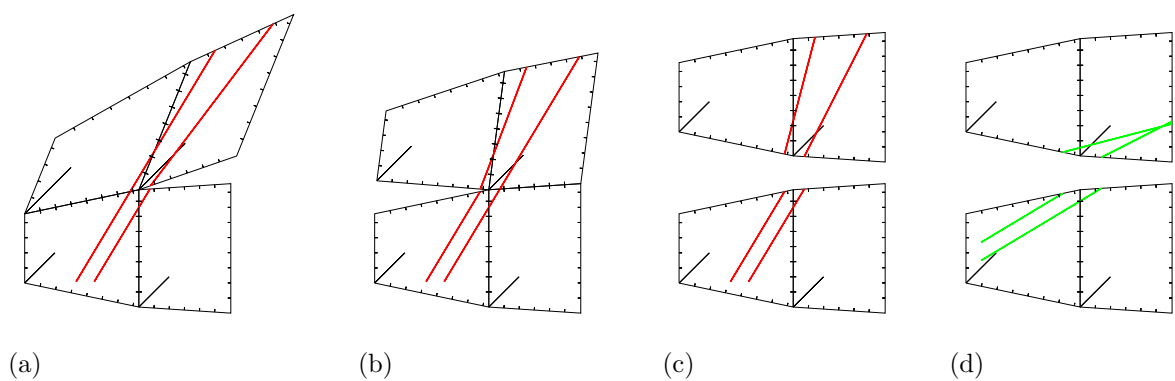

  \centering
  \subfigure[]{%
    \includegraphics[scale=\figelfscale]{\bilder/figure11a.eps}
  }\hfill%
  \subfigure[]{%
    \includegraphics[scale=\figelfscale]{\bilder/figure11b.eps}
  }\hfill%
  \subfigure[]{%
    \includegraphics[scale=\figelfscale]{\bilder/figure11c.eps}
  }\hfill%
  \subfigure[]{%
    \includegraphics[scale=\figelfscale]{\bilder/figure11d.eps}
  }
\caption{\label{fig.vertexgezeiten}
Anf\"anglich parallele Geod\"aten,
die an einem Vertex auf verschiedenen Seiten vorbeilaufen,
sind anschließend nicht länger parallel.
Die Sektoren sind in (a) längs der linken Geodäte und in (b) längs
der rechten Geodäte aneinandergelegt; in (c) sind sie symmetrisch
angeordnet.
(d) Raumartige Geodäten zeigen ein gegenläufiges Verhalten.
(Sektormodell wie in
Abb.~\protect\ref{fig.rtmodell2},
aber mit der doppelten zeitlichen Koordinatenlänge,
$c\Delta t = 2,5\,\rs$.)
}
\end{figure}

\subsection{
Konstruktion von Geodäten mit Transfer-Doppelsektoren}
\label{sec.trgeodzeichnen}

\begin{figure}
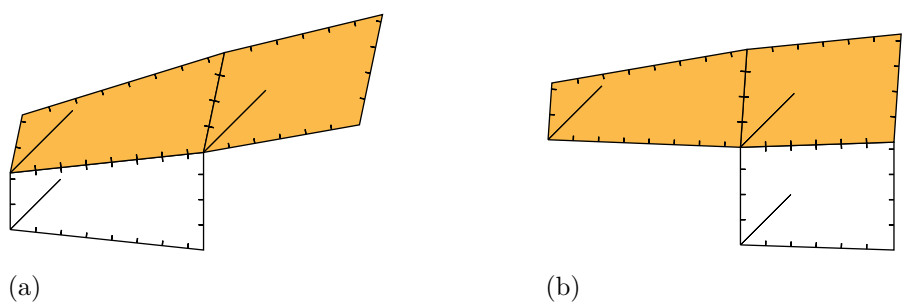

  \centering
  \subfigure[]{%
    \includegraphics[scale=0.5]{\bilder/figure12a.eps}
  }\hspace*{2cm}
  \subfigure[]{%
    \includegraphics[scale=0.5]{\bilder/figure12b.eps}
  }
  \caption{\label{fig.trtransfer}
Doppelsektoren als Transfersektoren.
(a) Für die linke Spalte des Sektormodells
von Abb.~\protect\ref{fig.rtmodell2},
(b) für die rechte Spalte.
}
\end{figure}

Bei der oben beschriebenen Konstruktion von Geodäten kommt es vor,
dass die Linie in einem angelegten Sektor sehr kurz ist,
weil sie nahe an einem Eckpunkt vorbeiläuft
(beispielsweise in Abb.~\ref{fig.gezeiten},
4. Zeile von unten, linke Linie).
In diesem Fall wird die Konstruktion sehr ungenau,
weil die weitere Richtung
aus dieser kurzen Linie bestimmt wird.
Das Problem lässt sich lösen, indem man statt eines einzelnen
Transfersektors einen Doppelsektor benutzt (Abb.~\ref{fig.trtransfer}).
Hierfür wird der Sektor der Nachbarspalte
nach passender Lorentztransformation an den Transfersektor angefügt.
Die Linie auf dem Doppelsektor ist dann länger,
so dass die Konstruktion genauer wird.
Im Workshop verwenden wir zur Einführung zunächst
die Einzelsektoren aus Abb.~\ref{fig.rtmodell2}.
Wenn das Prinzip klar geworden ist,
gehen wir zu den Doppelsektoren aus Abb.~\ref{fig.trtransfer} über.

\section{Fazit und Ausblick}

\label{sec.fazit3}

\subsection{Zusammenfassung und didaktische Anmerkungen}

Wir haben in diesem Beitrag gezeigt,
wie auf raumzeitlichen Sektormodellen
die Bahnen von Licht und freien Teilchen konstruiert werden können.
Die Konstruktion von Nullgeodäten führt auf direktem Weg
zum Phänomen der Gravitationsrotverschiebung
(Abschnitt~\ref{sec.rot}).
Die Konstruktion zeitartiger Geodäten verdeutlicht,
dass die Beschreibung eines nach oben geworfenen Teilchens
durch eine Geodäte in der Raumzeit
die erwartete Auf-/Abbewegung im Raum ergibt
(Abschnitt~\ref{sec.wurf}).
Anhand zeitartiger Geodäten benachbarter Teilchen
wird der Zusammenhang der raumzeitlichen Krümmung
mit der Newtonschen Gezeitenkraft deutlich
(Abschnitt~\ref{sec.gezeiten}).

Im Zusammenhang mit dem Einsatz raumzeitlicher Sektormodelle
kann das Äquivalenzprinzip thematisiert werden,
das hier deutlich zum Ausdruck kommt.
Es
besagt, dass
in genügend kleinen Gebieten einer gekrümmten Raumzeit
Minkowskigeometrie besteht und
lokal alle physikalischen Phänomene durch
die Spezielle Relativitätstheorie beschrieben werden.
Im Sektormodell stellt jeder einzelne Sektor ein solches
kleines Gebiet dar.
Die gekrümmte Raumzeit
wird explizit aus lokalen Bereichen mit Minkowskigeometrie
zusammengesetzt.
Im Sektormodell kann man sich von einem
Minkowskisektor zum nächsten
durch die gekrümmte Raumzeit hangeln.
Dass lokal die Spezielle Relativitätstheorie
gilt, wird
auf dem Sektormodell in direkter Weise umgesetzt,
wenn die Weltlinien von Licht und freien Teilchen
innerhalb eines Sektors als Geraden gezeichnet werden.

Das Sektormodell in diesem Beitrag stellt einen
1+1-dimensionalen Unterraum der Schwarzschild-Raumzeit dar.
Damit können
Weltlinien in radialer Richtung konstruiert werden.
Nichtradiale Weltlinien kann man im Prinzip in einem 2+1-dimensionalen
Sektormodell bestimmen, die Durchführung mit einem Papier- oder Pappmodell
erscheint jedoch nicht praktikabel. Eine Umsetzung
als dreidimensionale interaktive Computervisualisierung
wird derzeit untersucht.

Die in diesem Beitrag
vorgestellten Workshops zu Rotverschiebung
und Teilchenbahnen wurden im Rahmen
einer Einführung in die Allgemeine Relativitätstheorie
für Lehramtsstudierende an der Universität Hildesheim
entwickelt und erprobt
\citep{zah2013, kra2018}.
Dieser Einführungskurs
nutzt den hier beschriebenen modellbasierten Zugang
inklusive der Berechnung von Sektormodellen aus der Metrik.
Die Berechnung wird
ausgehend von der Kugeloberfläche über die Äquatorfläche
des Schwarzen Lochs (Teil~II, Abschnitt 2.4)
bis hin zur Raumzeit eines radialen Strahls
(Abschnitt~\ref{sec.trberechnung})
schrittweise eingeführt.
Der Kurs nutzt das in den Teilen~I bis~III vorgestellte
Material sowie weiteres Material aus dem in Vorbereitung
befindlichen vierten Teil.
In den Übungen zur Vorlesung
erstellen die Studierenden u.~a.
Sektormodelle zu weiteren Metriken
und untersuchen mit ihnen
Krümmung und Geodäten.
Der modellbasierte Kurs vermittelt also die Fähigkeit,
in einem gewissen Rahmen
anhand einer vorgegebenen Metrik physikalische Phänomene zu untersuchen.
Es werden dabei Aufgaben zeichnerisch gelöst, die in einem
Standardkurs rechnerisch bearbeitet würden.
Ein Beispiel für eine Aufgabe, die mit den Mitteln des
modellbasierten Kurses gelöst werden kann:
\glqq Gegeben ist die Metrik eines radialen Strahls in einer
expandierenden Raumzeit als
$\ud s^2 = -c^2 \ud t^2 + (t/T_0)^2\, \ud x^2$,
wobei $T_0$ eine Konstante ist.
Zwei Beobachter auf
festen Ortskoordinaten senden einander Lichtsignale zu.
Tritt eine Rotverschiebung auf?\grqq.
Über den Lehramtskurs und seine Evaluation
berichten \cite{kra2018}.

Weitere Einsatzmöglichkeiten, beispielsweise
in einer schulischen Arbeitsgemeinschaft Astronomie,
gibt es insbesondere für den Workshop
über Rotverschiebung, da er ohne Vorkenntnisse der Teilnehmer/innen
in Spezieller Relativitätstheorie auskommt.
Das gesamte Material kann auch als Ergänzung zu einer mathematisch
ausgerichteten Vorlesung eingesetzt werden
um das geometrische Verständnis zu fördern.

\subsection{Vergleich mit anderen anschaulichen Zugängen}

Sektormodelle ermöglichen eine bildliche Darstellung
raumzeitlicher Geodäten.
Andere bildliche Darstellungen von Geodäten in einer Raumzeit
wurden auf Einbettungsflächen
beschrieben
\citep{mar1999, jon2001, jon2005}.
Wie bei den hier vorgestellten Sektormodellen
sind auch diese Darstellungen auf 1+1-dimensionale Raumzeiten
beschränkt.
Mit der Darstellung auf Sektormodellen verwandt
ist die von diSessa entwickelte Konstruktion auf
sogenannten Keilkarten (wedge maps, \citealp{dis1981}).
Diese Konstruktion
ist vom Regge-Kalkül abgeleitet und
wird
numerisch durchgeführt.
Die Rechnung wird auch für 2+1-dimensionale Raumzeiten
beschrieben; Lichtablenkung und Rotverschiebung
werden diskutiert.

Im Vergleich zu Einbettungsflächen und auch zu
den Keilkarten
ist
die Berechnung und Nutzung von Sektormodellen elementarer.
Selbst für ein raumzeitliches Modell werden nur Grundkenntnisse
der Speziellen Relativitätstheorie benötigt;
die Bestimmung von Geodäten erfolgt zeichnerisch,
das einzige mathematische Konzept, das über Schulmathematik
hinausgeht, ist der Begriff der Metrik.
Sektormodelle lassen sich auf einfache Weise
berechnen
und
da sie sich
gut vervielfältigen lassen,
können alle Teilnehmer/innen eines Kurses die Konstruktion
von Geodäten auf eigenen Vorlagen selbst durchführen.

\subsection{Ausblick}

Die in Teil~I aufgeworfenen drei Grundfragen
nach dem Wesen einer gekrümmten Raumzeit,
den Gesetzmäßigkeiten für die Bewegung von Materie
und dem Zusammenhang zwischen der
Materieverteilung und der Krümmung der Raumzeit
sollen in dem hier vorgestellten modellbasierten Zugang
anhand von Sektormodellen behandelt werden.
In Teil~I wurden gekrümmte Räume und Raumzeiten im Modell dargestellt,
in Teil~II und dem vorliegenden Beitrag Geodäten als Bahnen
von Licht und freien Teilchen bestimmt.
Ein Folgebeitrag wird als vierter Teil den Zusammenhang von Krümmung und
Materieverteilung beschreiben.